\documentclass[12pt]{article}

\usepackage{amsmath}
\usepackage{amssymb}
\usepackage{shuffle}
\usepackage{physics}
\usepackage{tensor}
\usepackage{xcolor}
\usepackage{mathtools}
\usepackage{tikz}
\usepackage{tkz-euclide}
\usepackage{subcaption}
\usepackage[left=1in,right=1in,top=1in,bottom=1in]{geometry}
\usepackage{cite}

\usepackage{titling}
\usepackage{setspace}

\definecolor{citelink}{rgb}{0.33, 0.42, 0.18}
\definecolor{sectionfootnotelink}{rgb}{0.55, 0.0, 0.0}
\definecolor{equationlink}{rgb}{0.0, 0.2, 0.6}
\definecolor{urllink}{rgb}{0.53,0.15,0.34}
\definecolor{footnotelink}{rgb}{0.18,0.31,0.31}
\definecolor{integration_contours}{rgb}{1.0,0.0,0.22}
\definecolor{graph_poles}{rgb}{0.13, 0.67, 1.0}

\usepackage[linktoc=page,colorlinks,citecolor=citelink,linkcolor=sectionfootnotelink,urlcolor=urllink]{hyperref}
\usepackage{cleveref}
\crefname{equation}{}{}
\crefname{figure}{figure}{figures}

\makeatletter
\AddToHook{cmd/appendix/before}{\def\cref@section@alias{appendix}\def\cref@subsection@alias{appendix}}
\creflabelformat{equation}{\textup{\hypersetup{linkcolor=equationlink,linkbordercolor=equationlink}(#2#1#3)}}
\makeatother

\newcommand{\dexpval}[1]{%
  \mathinner{\langle\mkern-4mu\langle #1 \rangle\mkern-4mu\rangle}
}
\DeclareMathOperator{\sgn}{sgn}
\DeclareMathOperator{\Li}{Li}

\numberwithin{equation}{section}

\allowdisplaybreaks[4]

\begin{document}
        \begin{titlingpage}
        \vspace*{3em}
        \onehalfspacing
        \begin{center}
            {\LARGE
            \bf Form Factors in AdS from the Worldsheet}
        \end{center}
        \singlespacing
        \vspace*{2em}
        \begin{center}
            \textbf{
            Luis F. Alday\textsuperscript{a}, Cl\'ement Virally\textsuperscript{a} and Xinan Zhou\textsuperscript{b}
            }
        \end{center}
        \vspace*{1em}
        \begin{center}
            \textsl{
            \textsuperscript{a} Mathematical Institute, University of Oxford, Andrew Wiles Building, Radcliffe Observatory Quarter, Woodstock Road, Oxford, OX2 6GG, U.K.\\
            \textsuperscript{b} Kavli Institute for Theoretical Sciences, University of Chinese Academy of Sciences, Beijing 100190, China\\
            }
            \href{mailto:alday@maths.ox.ac.uk,}{\small alday@maths.ox.ac.uk}, \href{mailto:clement.virally@maths.ox.ac.uk}{\small clement.virally@maths.ox.ac.uk}, \href{mailto:xinan.zhou@ucas.ac.cn}{\small xinan.zhou@ucas.ac.cn}
        \end{center}
        \vspace*{3em}
        \begin{abstract}
            We study AdS form factors describing closed strings scattering off D-branes. Assuming that these form factors admit worldsheet integral representations in terms of multiple polylogarithms, we determine their leading curvature corrections around the flat-space limit by matching their pole structure to CFT data. We focus on two examples of defects in 4d $\mathcal{N}=4$ SYM: a $\frac{1}{2}$-BPS 't Hooft loop, dual to a D1 brane, and a pair of giant gravitons, dual to a D3 brane. For the 't Hooft loop, the worldsheet bootstrap completely fixes the form factor, and the result passes nontrivial checks from supersymmetric localization. For the giant gravitons, the bootstrap initially leaves five free coefficients, one of which can be fixed using localization. Remarkably, imposing instead that no new operators appear in the defect channel at this order fixes all five coefficients and reproduces the localization constraint. Under this additional assumption, we therefore obtain a unique D3-brane form factor, with supersymmetric localization providing an independent check.
            \end{abstract}
    \end{titlingpage}

    \tableofcontents

    \section{Introduction}
    Despite substantial progress, direct computations for strings scattering in AdS remain frustratingly out of reach in most cases. In flat space, compact expressions are available for tree-level superstring amplitudes at arbitrary multiplicity~\cite{Mafra:2011nv,Stieberger:2013wea}. In AdS, exact worldsheet methods are available in special backgrounds supported by the pure NS-NS flux~\cite{Maldacena:2001km}, but the situation is much more difficult in the presence of RR fluxes. In particular, the RNS formalism~\cite{PolchinskiString2} cannot be directly applied, while other formulations, such as the Green-Schwarz~\cite{Green:1980zg} and pure-spinor formalisms~\cite{Berkovits:2000fe}, have not yet been developed into practical tools for computing string scattering amplitudes in backgrounds such as AdS$_5\times$S$^5$.

    An alternative approach is to exploit the AdS/CFT correspondence~\cite{Maldacena:1997re,Gubser:1998bc,Witten:1998qj} together with the bootstrap philosophy. This strategy was first developed in the supergravity limit, where symmetry and consistency conditions can be powerful enough to determine tree-level holographic correlators without a direct computation from the bulk action \cite{Rastelli:2016nze,Rastelli:2017udc}. More recently, these ideas have been extended beyond supergravity and towards finite string tension. A key additional ingredient is the assumption of an emergent worldsheet representation for the AdS amplitude. Combined with CFT data and consistency conditions, this has made it possible to determine curvature corrections around the flat-space limit for both closed-string~\cite{Alday:2022uxp,Alday:2022xwz,Alday:2023jdk,Alday:2023mvu,Fardelli:2023fyq,Alday:2024rjs,Chester:2024wnb,Chester:2024esn,Wang:2025pjo,Jiang:2025oar} and open-string~\cite{Alday:2024yax,Alday:2024ksp,Wang:2025owf} amplitudes. In these constructions, the worldsheet integrands are built from multiple polylogarithms (MPLs)\footnote{More precisely, closed string amplitudes are built out of single-valued multiple polylogarithms (SVMPLs)~\cite{Alday:2023jdk}, while open strings are built out of the single-variable, non-single-valued version~\cite{Alday:2024yax,Alday:2024ksp}. For a review of these functions, see \cref{sec:mpls}.}, and exhibit remarkable structures such as KLT and monodromy relations~\cite{Alday:2025bjp,Alday:2025cxr,Nocchi:2026yms}.

    In this paper, we extend this worldsheet bootstrap approach to AdS form factors describing closed strings scattering off D-branes. Through AdS/CFT, these observables are described by two-point functions of local operators in the presence of conformal defects. In flat space, string scattering from a D-brane is described by a disk amplitude with a universal Veneziano-like structure~\cite{Klebanov:1995ni,Hashimoto:1996bf,Garousi:1996ad}
    \begin{equation}
        \mathcal{A}^{(0)}(S,Q) = \int_0^1 \dd x\, x^{-\frac{S}{2}-1}(1-x)^{-Q-1}(-a_1(1-x)+a_2 x) = (2 a_1 Q-a_2 S)\frac{\Gamma(-Q)\Gamma\left(-\frac{S}{2}\right)}{2\Gamma\left(1-Q-\frac{S}{2}\right)}\;,\label{eq:flat_space_form_factor}
    \end{equation}
    where $a_1$ and $a_2$ are kinematic quantities. Our goal is to determine the leading curvature correction to this flat-space form factor while retaining its full dependence on the string-scale kinematic variables. At large AdS radius, the form factor admits a curvature expansion of the form
    \begin{equation}
        \mathcal{A}(S,Q) = \mathcal{A}^{(0)}(S,Q) + \frac{\alpha'}{R^2}\mathcal{A}^{(1)}(S,Q)+\frac{(\alpha')^2}{R^4}\mathcal{A}^{(2)}(S,Q)+\ldots
    \end{equation}
    We focus in this paper on the first correction $\mathcal{A}^{(1)}(S,Q)$. Motivated by the worldsheet representations found for AdS scattering amplitudes, we assume that it also admits an analogous worldsheet integral representation
    \begin{equation}
        \mathcal{A}^{(1)}(S,Q) = \int_0^1\dd x\, x^{-\frac{S}{2}-1}(1-x)^{-Q-1}\mathcal{G}^{(1)}(S,Q;x)\;,\label{eq:ws_flat_general}
    \end{equation}
    where $\mathcal{G}^{(1)}(S,Q;x)$ is built from multiple polylogarithms. On the CFT side, the same observable is encoded in a defect two-point function. Its spectrum and OPE coefficients determine the poles and residues of the Mellin amplitude, whose flat-space limit gives the corresponding string form factor~\cite{Alday:2024srr,Chen:2025cod}. In \cite{Alday:2024srr}, this relation was used to determine stringy corrections order by order in a low-energy, or higher-derivative, expansion around supergravity. Here we instead aim to determine the complete first curvature correction, retaining its full dependence on $\alpha'S$ and $\alpha'Q$. To this end, we match the pole structure and residues implied by the defect CFT to those of the worldsheet representation \cref{eq:ws_flat_general}.

    We apply this strategy to two examples in 4d $\mathcal{N}=4$ SYM: a $\frac{1}{2}$-BPS 't Hooft line, dual to a D1 brane in AdS$_5\times$S$^5$, and a pair of maximal giant gravitons, which can be viewed as a zero-dimensional defect and are dual to a D3 brane~\cite{Chen:2025yxg,Chen:2026ium}. In both cases, matching the poles and residues of the worldsheet representation to the defect CFT data strongly constrains the leading curvature correction. For the 't Hooft line, the worldsheet bootstrap uniquely fixes the form factor, including the leading Regge trajectory, and the result passes nontrivial independent checks from supersymmetric localization~\cite{Pufu:2023vwo,Billo:2023ncz,Dempsey:2024vkf,Billo:2024kri}. For the giant gravitons, we use the known leading Regge trajectory~\cite{Gromov:2011de,Basso:2011rs,Gromov:2011bz} as an additional input, after which the worldsheet bootstrap leaves five coefficients unfixed. Supersymmetric localization~\cite{Brown:2024tru} fixes one of these coefficients. Alternatively, requiring that no new operators appear in the defect channel at this order fixes all five coefficients without using localization, and the resulting form factor reproduces the localization constraint. Similar behaviour, that no new operators appear in curvature corrections, had previously been observed in the defect-free case~\cite{Alday:2025pmg}. Motivated by the uniqueness of this solution and its agreement with localization, we conjecture that it gives the correct giant-graviton form factor.

    The rest of this paper is organized as follows. In \cref{sec:setup}, we present our setup in both theories, and the Mellin approach to defect CFT. In \cref{sec:approaches}, we give more details on the flat-space limit of Mellin amplitudes on the one hand, and on our worldsheet ansatz on the other. In \cref{sec:results}, we present our final results for the worldsheet integrand for the form factors, as well as some new CFT data obtained by our methods, and compare our results to supersymmetric localization. Finally, we conclude in \cref{sec:discussion}. Several appendices contain various technical points.

    \section{Setup}\label{sec:setup}
    \subsection{Line defects in \texorpdfstring{$\mathcal{N}=4$}{N=4} SYM}\label{sec:line_defect_setup}
    The first case of defects we consider is a $\frac{1}{2}$-BPS 't Hooft line in $\mathcal{N}=4$ SYM, which is dual to a D1-brane in the AdS. We consider the correlation function of two bulk local operators in the presence of such a line defect. More precisely, we will focus on the superconformal primary of the stress tensor multiplet
    \begin{equation}
            \mathcal{O}_2(x,t) = \mathcal{N}_2\tr\left(\Phi^I(x)t^I\right)^2\;,
        \end{equation}
    which has protected conformal dimension $\Delta=2$ and transforms in the rank-$2$ symmetric traceless representation of the $SO(6)$ R-symmetry group. Here $\Phi^I$ with $I=1,\ldots,6$ are the six scalars of $\mathcal{N}=4$ SYM. The null vector $t^I$, satisfying $t\cdot t=0$, contracts the R-symmetry indices and projects the operator to the symmetric traceless representation. We also choose the normalization $\mathcal{N}_2$ such that the two-point function in the absence of the defect is unit normalized
    \begin{equation}
        \langle \mathcal{O}_2(x_1,t_1) \mathcal{O}_2(x_2,t_2)\rangle=\frac{(t_1\cdot t_2)^2}{x_{12}^4}\;.
    \end{equation}
    In the presence of the $\frac{1}{2}$-BPS 't Hooft line, the two-point function  can be written as\footnote{In fact, this applies to general $\frac{1}{2}$-BPS line defects, including in particular the Wilson line.}\footnote{We denote correlators in the presence of defects by double angle brackets. To be completely explicit, if we denote the line defect as $\mathbb{L}$, the defect correlator is defined by $\dexpval{\mathcal{O}\mathcal{O}}\equiv\frac{\expval{\mathcal{O}\mathcal{O}\mathbb{L}}}{\expval{\mathbb{L}}}$.}
    \begin{equation}\label{linedefect2ptfun}      \dexpval{\mathcal{O}_1(x_1,t_1)\mathcal{O}_2(x_2,t_2)} = \frac{(t_1\cdot\theta)^2(t_2\cdot\theta)^2}{|x_1^i|^2|x_2^i|^2}\mathcal{F}(\zeta,\eta;\sigma)\;,
    \end{equation}
    where $i=2,\ldots,4$ labels the coordinates transverse to the line defect, and the direction $x^1$ is along the defect and is denoted as $\tau$. The 't Hooft line is also associated with a fixed unit vector $\theta$ which breaks the $SO(6)$ R-symmetry to $SO(5)$. Holographically, the 't Hooft line corresponds to AdS$_2$ surface in AdS$_5$, and $\theta$ corresponds to the point on S$^5$ where the defect is located. In \cref{linedefect2ptfun} we have also defined the conformal and R-symmetry cross-ratios
    \begin{equation}
        \zeta = \frac{(\tau_1-\tau_2)^2+|x_1^i|^2+|x_2^i|^2}{2|x_1^i||x_2^i|}\;,\quad \eta = \frac{x_1^j x_2^j}{|x_1^i||x_2^i|}\;,\quad \sigma = \frac{(t_1\cdot t_2)}{(t_1\cdot \theta)(t_2\cdot \theta)}\;.
    \end{equation}
    As a function of $\sigma$, $\mathcal{F}(\zeta,\eta;\sigma)$ is a quadratic polynomial, which we can explicitly write as
    \begin{equation}
        \mathcal{F}(\zeta,\eta;\sigma) = \mathcal{F}_0(\zeta,\eta) + \sigma \mathcal{F}_1(\zeta,\eta) + \sigma^2\mathcal{F}_2(\zeta,\eta)\;.\label{eq:F_sigma_decomposition}
    \end{equation}
    This accounts for the invariance under the bosonic generators of the superconformal group, but we must also consider the fermionic generators. Invariance under their action requires the correlator to obey the superconformal Ward identities (SCWI), which take the form~\cite{Liendo:2016ymz}:
    \begin{equation}
    \Bigg(\partial_z+\frac{1}{2}\partial_\alpha\Bigg)\eval{\mathcal{F}(z,\bar z;\alpha)}_{\alpha=z}=0\;,\label{eq:scwi}
    \end{equation}
    plus a second one with $z\leftrightarrow\bar z$. Written in terms of the $\mathcal{F}_i(\zeta,\eta)$, the SCWI become a set of coupled differential equations for three functions. The $(z,\bar z,\alpha)$ coordinates are related to $(\zeta,\eta,\sigma)$ by
    \begin{equation}
        \zeta = \frac{1+z\bar z}{2\sqrt{z\bar z}}\;,\quad \eta = \frac{z+\bar z}{2\sqrt{z\bar z}}\;,\quad \sigma = -\frac{(1-\alpha)^2}{2\alpha}\;.
    \end{equation}
    Physically, we can understand the $z,\bar z$ as follows. Using conformal symmetry, we can always choose to put the two operators on a two-dimensional plane, with the following coordinates:
    \begin{equation}
        x_1 = \left(\frac{z-\bar z}{2i},\frac{z+\bar z}{2},0,0\right),\quad x_2 = (0,1,0,0)\;.
    \end{equation}
    We can also align the defect with the line defined by $x^2=x^3=x^4=0$. As a consequence of the SCWI \cref{eq:scwi}, the twisted correlator $\mathcal{F}(z,z;z)$ is just a number, which we denote by $\mathcal{F}_{\text{top}}(\sgn z)$. As the subscripts denotes, this is almost topological, but depends on the sign of $z$, which is to say that $\mathcal{F}_{\text{top}}(\sgn z)$ is different based on whether the two operators are inserted on the same side of the defect in this two-dimensional plane, or on opposite sides.

    Let us now introduce the Mellin representation of defect correlators \cite{Rastelli:2017ecj,Goncalves:2018fwx}. This representation, which shares many similarities with the defect-free case, can be viewed as representing holographic defect correlators in AdS ``momentum space''. It defines an AdS form factor of particles scattering off of an extended object. It is more convenient to introduce a different combination of cross-ratios 
    \begin{equation}\label{zetaetaxichi}
        \zeta = \frac{\xi+\chi}{2},\quad \eta = \frac{\chi}{2}\;.
    \end{equation}
    Then the defect two-point function can be written as 
    \begin{equation}
        \mathcal{F}(\xi,\chi;\sigma) = \int \frac{\dd\gamma\,\dd\delta}{(2\pi i)^2}\,\xi^{-\delta}\chi^{-\gamma+\delta}\mathcal{M}(\delta,\gamma;\sigma)\Gamma(\delta)\Gamma(\gamma-\delta)\prod_{i=1}^2\Gamma\left(\frac{\Delta_i-\gamma}{2}\right)\;,\label{eq:mellin_transform}
    \end{equation}
    where the integration contours are parallel to the imaginary axis and for the stress tensor multiplet we should set $\Delta_i=2$. The nontrivial dynamics are then encoded in the Mellin amplitude $\mathcal{M}(\delta,\gamma;\sigma)$. As a function of $\sigma$, the Mellin amplitude inherits the quadratic polynomial structure of $\mathcal{F}(\xi,\chi;\sigma)$, allowing us to write it as
    \begin{equation}
        \mathcal{M}(\delta,\gamma;\sigma) = \mathcal{M}_0(\delta,\gamma) + \sigma\mathcal{M}_1(\delta,\gamma) + \sigma^2\mathcal{M}_2(\delta,\gamma)\;.\label{eq:mellin_sigma_decomposition}
    \end{equation}
    In~\cite{Gimenez-Grau:2023fcy}, it was shown that the SCWI can be written in Mellin space as a set of coupled difference equations among the $\mathcal{M}_i(\delta,\gamma)$, similar to the situation in the defect-free case~\cite{Zhou:2017zaw}. In~\cite{Virally:2025nnl}, it was shown that there exists a general solution to the SCWI in Mellin space in terms of a difference operator $\mathcal{D}$, which depends on $\sigma$, acting on a single unconstrained function $\widetilde{\mathcal{M}}(\delta,\gamma)$ with no $\sigma$ dependence, which we call the reduced Mellin amplitude: 
    \begin{equation}
        \mathcal{M}(\delta,\gamma;\sigma) = \mathcal{D}\mathcal{\widetilde{M}}(\delta,\gamma)\;.\label{eq:reduced_definition}
    \end{equation}
    We review the SCWI in Mellin space, as well as their solution, in \cref{sec:scwi_mellin}. We note here that, as in \cref{eq:mellin_sigma_decomposition}, the different powers of $\sigma$ separate when we write the solution like this, such that we need only know one of the three to obtain the reduced amplitude.

    Finally, we will need the Mellin amplitude in the supergravity limit, which encodes the contribution of massless states to the AdS form factor in the flat-space limit. In \cite{Gimenez-Grau:2023fcy}, it was found to be:
    \begin{equation}
        \begin{aligned}
        \mathcal{M}_{\text{line,SUGRA}}(\delta,\gamma;\sigma) \propto \Bigg({}&
        \frac{\sigma\left(\sigma(\gamma-\delta)-\sigma+4\right)}{\delta-1}-\frac{2(\gamma-\delta)+4\sigma-4}{\gamma}+\frac{\gamma-\delta}{\gamma+2}+(\sigma-1)^2\\
        &+\frac{2^{\gamma+2}\Gamma(-\gamma)\left(\frac{\delta+2}{\gamma+2}-\sigma\right)}{\Gamma\!\left(\frac{2-\gamma}{2}\right)^2}
        \Bigg)\;.\label{eq:line_sugra}
        \end{aligned}
    \end{equation}
    Here we have not fixed the normalization which will later be determined by requiring the amplitude to be compatible with the flat space result \cref{eq:flat_space_form_factor}. In~\cite{Virally:2025nnl}, it was shown that this has a reduced Mellin amplitude:
    \begin{equation}
        \widetilde{\mathcal{M}}_{\text{line,SUGRA}}(\delta,\gamma) = \frac{2}{\gamma-2} + \frac{(\gamma-3)2^\gamma\Gamma(2-\gamma)}{(\gamma-4)\Gamma\left(2-\frac{\gamma}{2}\right)^2}\;,\label{eq:line_sugra_reduced}
    \end{equation}
    which has the surprising property that it is independent of $\delta$.

    \subsubsection{Exchanges in two channels}\label{sec:two_channels}
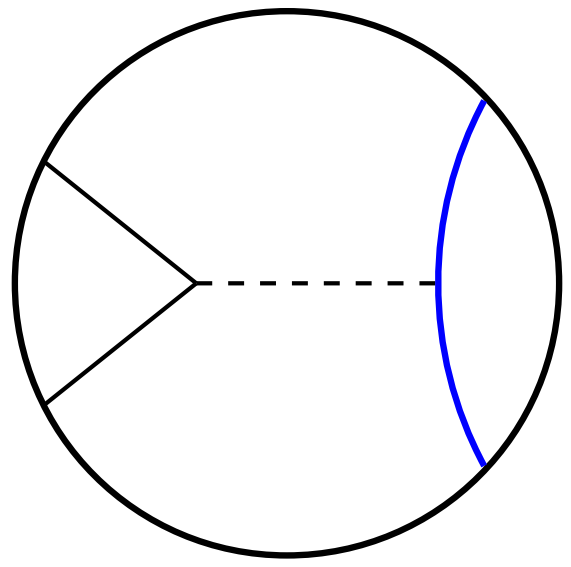
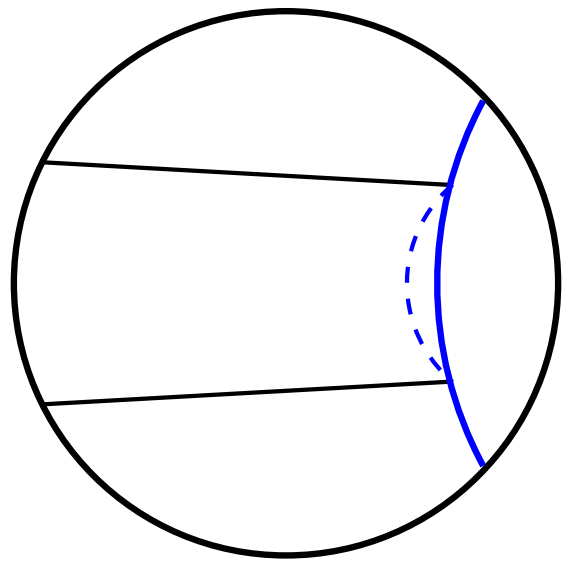
\begin{figure}
    \centering
    \begin{subfigure}{0.45\linewidth}
        \centering
        \begin{tikzpicture}
            \draw[very thick] (0,0) circle (1.8);

            \draw[thick] (-1.6,0.8) -- (-0.6,0);
            \draw[thick] (-1.6,-0.8) -- (-0.6,0);

            \draw[thick,dashed] (-0.6,0) -- (1,0);

            \tkzDefPoint(1.3,1.2){A}
            \tkzDefPoint(1,0){B}
            \tkzDefPoint(1.3,-1.2){C}
            \tkzCircumCenter(A,B,C)
            \tkzGetPoint{O}
            \tkzDrawArc[delta=0.2,color=blue,very thick](O,A)(C)
        \end{tikzpicture}
        \subcaption{Bulk channel exchange}
    \end{subfigure}
    \begin{subfigure}{0.45\linewidth}
        \centering
        \begin{tikzpicture}
            \draw[very thick] (0,0) circle (1.8);

            \draw[thick] (-1.6,0.8) -- (1.1,0.65);
            \draw[thick] (-1.6,-0.8) -- (1.1,-0.65);

            \tkzDefPoint(1.1,0.65){D}
            \tkzDefPoint(0.8,0){E}
            \tkzDefPoint(1.1,-0.65){F}
            \tkzCircumCenter(D,E,F)
            \tkzGetPoint{Q}
            \tkzDrawArc[delta=0.2,color=blue,thick,dashed](Q,D)(F)
            
            \tkzDefPoint(1.3,1.2){A}
            \tkzDefPoint(1,0){B}
            \tkzDefPoint(1.3,-1.2){C}
            \tkzCircumCenter(A,B,C)
            \tkzGetPoint{O}
            \tkzDrawArc[delta=0.2,color=blue,very thick](O,A)(C)
        \end{tikzpicture}
        \subcaption{Defect channel exchange}
    \end{subfigure}
    \caption{The different exchange channels. Note the similarity of the bulk channel to the standard $s$ channel exchanges, while the defect channel resembles the $t$ and $u$ channels. Worldlines in the bulk are shown in black, while those on the the defect are shown in blue. Dashed lines denote exchanged particles, while the full black lines are particles which couple to local operators in the CFT. The full blue line denotes the AdS$_{p+1}$ subspace of AdS$_{d+1}$ which is dual to the defect.}\label{fig:witten_diagrams}
\end{figure}

    In Mellin space, we can decompose a Mellin amplitude $\mathcal{M}(\delta,\gamma)$%
    \footnote{For this section, we will ignore SUSY and thus the dependence on $\sigma$, and focus on bosonic exchanges. We will return to the constraints imposed by SUSY in \cref{sec:superconformal_exchanges}.}
    into sums of basic kinematic quantities $M_{\Delta,\ell}(\delta,\gamma)$, which have the right poles to recreate the conformal blocks in position space:
    \begin{equation}
        \mathcal{M}(\delta,\gamma) = \sum_{\mathcal{O}}\lambda_{12\mathcal{O}}\,M_{\Delta,\ell}(\delta,\gamma)\;.
    \end{equation}
    Here $\Delta,\ell$ are respectively the conformal dimension and spin of the primary operator $\mathcal{O}$, and $\lambda_{12\mathcal{O}}$ is the appropriate product of OPE coefficients for the exchange of $\mathcal{O}$. There are two kinds of exchanges we may consider in the presence of a defect, which depend on where the exchanged operator lives. These are analogous to the $s,t,u$ channels of the defect-free case, in the sense that they correspond to different forms of Witten diagrams, as illustrated in \cref{fig:witten_diagrams}.\footnote{Strictly speaking, the kinematic building blocks are the conformal blocks. The exchange Witten diagram contains such a single-trace conformal block, but also infinitely many double-trace operators formed out of the external operators. But as we will point out they have the same poles and residues in the Mellin amplitude. Therefore, we will slightly abuse the terminologies by using them interchangeably.}

    In the first case, known as the bulk channel, the local operators meet in the bulk and exchange an operator, which lives in the bulk, with the defect. The second case, known as the defect channel, involves the local operators approaching to the defect and exchanging an operator which lives on the defect.
    
    In Mellin space, conformal blocks have the same poles and residues as the exchange Witten diagrams. Taking the residues at these poles reproduces the power series of conformal blocks. For the case of defect two-point functions, the Mellin amplitudes have poles either in $\delta$ or in $\gamma$. The poles in $\delta$ are associated with the exchanges in the bulk channel, while the exchanges in the defect channel have poles in $\gamma$~\cite{Gimenez-Grau:2023fcy}:
    \begin{equation}
        M_{\Delta,\ell}^{\Delta_1,\Delta_2,\text{bulk}}(\delta,\gamma) = \sum_{n=0}^\infty \frac{Q_{\Delta,\ell,n}^{\Delta_1,\Delta_2}(\gamma)}{\delta-\frac{\Delta_1+\Delta_2-\tau-2n}{2}},\quad M_{\Delta,s}^{\Delta_1,\Delta_2,\text{def}}(\delta,\gamma) = \sum_{n=0}^\infty\frac{P_{\Delta,s,n}^{\Delta_1,\Delta_2}(\delta,\gamma)}{\gamma-\tau-2n}\;.\label{eq:mellin_poles}
    \end{equation}
    Here $\Delta$ is the conformal dimension, $\ell,s$ are respectively the spin and transverse spin of the bulk and defect operator. We have also defined the conformal twist $\tau$ which is either $\tau=\Delta-\ell$ or $\tau=\Delta-s$. In the numerators $Q_{\Delta,\ell,n}^{\Delta_1,\Delta_2}(\gamma)$ and $P_{\Delta,s,n}^{\Delta_1,\Delta_2}(\delta,\gamma)$ are defect analogues of the Mack polynomials~\cite{Mack:2009mi}. $Q_{\Delta,\ell,n}^{\Delta_1,\Delta_2}(\gamma)$ is a polynomial of degree $\frac{\ell}{2}$, while $P_{\Delta,s,n}^{\Delta_1,\Delta_2}(\delta,\gamma)$ has degree $s$. We show how to construct the Mack polynomials explicitly in \cref{sec:conformal_blocks_mellin}. In all relevant cases for our work, the bulk channel exchanges only even spin operators, while the defect channel allows for both odd and even transverse spins.

    \subsubsection{Superconformal exchanges in \texorpdfstring{$\mathcal{N}=4$}{N=4} SYM with line defects}\label{sec:superconformal_exchanges}
    We must now understand the consequences of SUSY on the exchange diagrams in the last section. The constraints of SUSY are encapsulated in the SCWI \cref{eq:scwi}. This equation combines the exchanges of individual bosonic operators, given in the last section, into exchanges of full superconformal multiplets. We can thus write an exchange which obeys the SCWI as:
    \begin{equation}
        \mathcal{M}(\delta,\gamma;\sigma) = \sum_{\mathcal{O}}\lambda_{12\mathcal{O}}\overline{M}_{\Delta,\ell}(\delta,\gamma;\sigma)\;,\label{eq:block_decomposition}
    \end{equation}
    where $\overline{M}_{\Delta,\ell}(\delta,\gamma;\sigma)$ is a superconformal block. We will now specialize to the case $\Delta_1=\Delta_2=2$, which is most relevant to our work, and suppress the $\Delta_i$ dependence from the bosonic blocks. We then write:
    \begin{equation}
        \overline{M}_{\Delta,\ell}(\delta,\gamma;\sigma) = \sum_{i,j,k}f_{ijk}\mathcal{R}_k(\sigma) M_{\Delta+i,\ell+j}(\delta,\gamma)\label{eq:superconformal_exchange_full}
    \end{equation}
    where $\Delta,\ell$ in the above equation are the conformal dimension and spin of a superconformal primary operator, $M_{\Delta,\ell}(\delta,\gamma)$ is a single bosonic exchange from the last section, $\mathcal{R}_k(\sigma)$ is an R-symmetry polynomial, which we will shortly define in each channel, and $f_{ijk}$ are numerical coefficients which depend on $\Delta,\ell$. The $f_{ijk}$ can be obtained by demanding that each of these superconformal exchange diagrams solves the SCWI. We refer readers interested in the full details of these superconformal blocks to~\cite{Liendo:2016ymz}. We will adopt a different strategy here. First, we note that the R-symmetry polynomials (as well as the $f_{ijk}$) are different in the bulk and defect channels:
    \begin{equation}
        \begin{aligned}
        \mathcal{R}_k^{\text{bulk}}(\sigma) ={}& \sigma^{\frac{\Delta_1+\Delta_2-k}{2}}\, _2F_1\left(-\frac{k}{2},-\frac{k}{2};-k-1;\frac{\sigma}{2}\right)\;,\\
        \quad \mathcal{R}_k^{\text{def}}(\sigma) ={}& (-2\sigma)^k\, _2F_1\left(-k-1,-k;-2(k+1);\frac{2}{\sigma}\right)\;.
        \end{aligned}
    \end{equation}
    In the bulk channel, R-symmetry polynomials have $k=0,2,4$ while in the defect channel, $k=0,1,2$. We note in particular that only $\mathcal{R}_4^{\text{bulk}}(\sigma)$ contains $\sigma^0$ and only $\mathcal{R}_2^{\text{def}}$ contains $\sigma^2$. It is also the case that each of these polynomials only show up once in \cref{eq:superconformal_exchange_full}. We can then choose the overall normalization of the superconformal blocks such that:
    \begin{equation}
        \begin{aligned}
            \eval{\overline{M}^{\text{bulk}}_{\Delta,\ell}(\delta,\gamma;\sigma)}_{\sigma^0} = M_{\Delta+4,\ell}^{\text{bulk}}(\delta,\gamma)\;,\quad \eval{\overline{M}^{\text{def}}_{\Delta,\ell}(\delta,\gamma;\sigma)}_{\sigma^2} = M_{\Delta+2,\ell}^{\text{def}}(\delta,\gamma)\;.\label{eq:superconformal_diagrams}
        \end{aligned}
    \end{equation}
    Then, \cref{eq:block_decomposition} becomes:
    \begin{equation}
        \mathcal{M}(\delta,\gamma;\sigma) = \sum_{\mathcal{O}_b}\lambda_{12\mathcal{O}_{b}}\,\overline{M}_{\Delta,\ell}^{\text{bulk}}(\delta,\gamma;\sigma) = \sum_{\mathcal{O}_d}\lambda_{12\mathcal{O}_d}\,\overline{M}_{\Delta,\ell}^{\text{def}}(\delta,\gamma;\sigma)\;,\label{eq:two_channels_line}
    \end{equation}
    where $\mathcal{O}_b$ is an operator in the bulk channel, and $\mathcal{O}_d$ is an operator in the defect channel, and we highlight that we should get the same answer in both channels. We can, in principle, use \cref{eq:superconformal_diagrams} to invert the SCWI solution \cref{eq:reduced_definition} and find a reduced form for the exchange diagram, which contains the information for the full superconformal block. This was shown to be possible for defect-free theories with eight supercharges in~\cite{Chester:2025jxg}. In practice, it is easiest to rewrite this in reduced form after taking the flat-space limit, as explained in \cref{sec:scwi_flat_limit}. We also point out that the shifts of $\Delta\to\Delta+4$ in the bulk channel and $\Delta\to\Delta+2$ in the defect channel are the consequence of the superconformal symmetry.

    \subsection{Giant gravitons}
We also consider a second type of scattering with defects. In this case, the defect is a pair of heavy operators in $\mathcal{N}=4$ SYM known as the giant graviton \cite{McGreevy:2000cw,Balasubramanian:2001nh,Hashimoto:2000zp}. While they can be regarded as ordinary local operators, in \cite{Chen:2025yxg,Chen:2026ium} it was shown that the most natural and computationally advantageous view point is to treat them as a special kind of zero dimensional defects. 

More precisely, the giant graviton (maximal in this case) is a $\frac{1}{2}$-BPS operator defined as a determinant in color space
\begin{equation}
        \mathcal{D}_{\text{GG}}(x,t) = \det\left(\Phi^I(x)t^I\right)\;.
    \end{equation}
It has protected conformal dimension $N$ and transforms in the rank-$N$ symmetric traceless representation. We consider the light-light-heavy-heavy four-point function, which can be written in the following form as a consequence of the superconformal Ward identities \cite{Eden:2000bk,Nirschl:2004pa}
\begin{equation}\label{OODD4pt}
\expval{\mathcal{O}_2(x_1,t_1)\mathcal{O}_2(x_2,t_2)\mathcal{D}_{\text{GG}}(x_3,t_3)\mathcal{D}_{\text{GG}}(x_4,t_4)} =G_{\text{free}}(x_i,t_i) + t_{12}^2t_{34}^2x_{13}^4x_{24}^4\mathcal{R}\mathcal{H}(U,V)\;,
    \end{equation}
Here $G_{\text{free}}(x_i,t_i)$ is the free part of the connected correlator and $\mathcal{H}(U,V)$ is the reduced correlator. We have defined the cross-ratios:
    \begin{equation}
        \begin{gathered}
            U = \frac{x_{12}^2x_{34}^2}{x_{13}^2x_{24}^2} = z\bar z\;,\quad V = \frac{x_{14}^2x_{23}^2}{x_{13}^2x_{24}^2}=(1-z)(1-\bar z)\;,\\
            \sigma = \frac{t_{13}t_{24}}{t_{12}t_{34}} = \alpha\bar\alpha\;,\quad \tau = \frac{t_{14}t_{23}}{t_{12}t_{34}} = (1-\alpha)(1-\bar\alpha)\;.
        \end{gathered}
    \end{equation}
    $\mathcal{R}$ is a factor determined by superconformal symmetry:
    \begin{equation}
        \mathcal{R} = (1-z\alpha)(1-z\bar\alpha)(1-\bar z\alpha)(1-\bar z\bar\alpha)\;,
    \end{equation}
which plays a similar role as the operator $\mathcal{D}$ in the line defect case. Note in particular the reduced correlator $\mathcal{H}(U,V)$ is an R-symmetry singlet and does not depend on $\sigma$ and $\tau$.
    
To switch to the defect picture, we simply divide the four-point function by the heavy two-point function
\begin{equation}
\dexpval{\mathcal{O}_2(x_1,t_1)\mathcal{O}_2(x_2,t_2)}\equiv \frac{\expval{\mathcal{O}_2(x_1,t_1)\mathcal{O}_2(x_2,t_2)\mathcal{D}_{\text{GG}}(x_3,t_3)\mathcal{D}_{\text{GG}}(x_4,t_4)}}{\expval{\mathcal{D}_{GG}(x_3,t_3)\mathcal{D}_{GG}(x_4,t_4)}}\;.
    \end{equation}
Then \cref{OODD4pt} becomes
    \begin{equation}
        \dexpval{\mathcal{O}_2(x_1,t_1)\mathcal{O}_2(x_2,t_2)} = F_{\text{free}}(x_i,t_i) + \frac{t_{12}^4x_{13}^4x_{24}^4}{x_{34}^4}\mathcal{R}\mathcal{F}(U,V)\;,
    \end{equation}
where $F_{\text{free}}$ and $\mathcal{F}(U,V)$ are respectively the free and reduced defect two-point functions. In this defect picture, it is more natural to introduce the defect cross-ratios. They are defined in terms of the projector\footnote{There is also another projector for R-symmetry which is constructed in terms of the polarization vectors $t$. But here because we focus on the reduced defect correlator, only the spacetime part is needed. As emphasized in \cite{Chen:2025yxg,Chen:2026ium}, an important feature is that the heavy operators only enter as projectors in the defect picture.}
\begin{equation}
        \mathbb{N}_{AB} = \frac{P_{3,A}P_{4,B} + P_{4,A}P_{3,B}}{P_3\cdot P_4}\;,
    \end{equation}
where we have used the embedding space and $P\in \mathbb{R}^{5,1}$ are six dimensional null vectors 
\begin{equation}
        P_A = \left(\frac{1+x^2}{2},\frac{1-x^2}{2},\vec{x}\right)\;.
    \end{equation}
The defect cross-ratios are
    \begin{equation}
        \xi = \frac{-2P_1\cdot P_2}{\left(\prod_{i=1}^2P_i\cdot\mathbb{N}\cdot P_i\right)^{\frac{1}{2}}} = \frac{U}{\sqrt{V}}\;,\quad \chi = \frac{2P_1\cdot P_2 - 2P_1\cdot\mathbb{N}\cdot P_2}{\left(\prod_{i=1}^2P_i\cdot\mathbb{N}\cdot P_i\right)^{\frac{1}{2}}} = \frac{1-U+V}{\sqrt{V}}\;,
    \end{equation}
    which should be identified with $\xi$ and $\chi$ introduced in \cref{zetaetaxichi}. We can also define the Mellin representation \cite{Chen:2026ium} 
    \begin{equation}
        \mathcal{F}(\xi,\chi) =\int \frac{\dd\delta\dd\gamma}{(2\pi i)^2}\xi^{-\delta}\chi^{\delta-\gamma-1}\mathcal{M}(\delta,\gamma)\Gamma(\delta)\Gamma(\gamma-\delta+1)\Gamma\left(\frac{3-\gamma}{2}\right)^2\;,
    \end{equation}
    where the Gamma function factor is different because we are working with the reduced correlator. Upon setting $\Delta_i=4$ and shifting $\gamma\to\gamma+1$, we can bring it to the same form as \cref{eq:mellin_transform} with $p=0$. Therefore, we identify:
    \begin{equation}
        \mathcal{M}_{\text{GG}}(\delta,\gamma) = \mathcal{M}^{p=0,\Delta_i=4}(\delta,\gamma+1)\;,
    \end{equation}
    where $\mathcal{M}^{p,\Delta_i}(\delta,\gamma)$ is the standard defect Mellin amplitude we presented in \cref{sec:two_channels}, for a defect of dimension $p$. Other than this shift of $\gamma$, these Mellin amplitudes are identical to the ones we presented for the line defects. In particular, their decomposition into blocks in both the bulk and defect channels is given by:
    \begin{equation}
        \mathcal{M}^{p=0,\Delta_i=4}(\delta,\gamma) = \sum_{\mathcal{O}_b} \lambda_{12\mathcal{O}_{b},\text{bulk}}M_{\Delta+4,\ell,\text{bulk}}^{p=0,\Delta_i=4}(\delta,\gamma) = \sum_{\mathcal{O}_d} \lambda_{12\mathcal{O}_{d},\text{def}}M_{\Delta+2,\ell,\text{def}}^{p=0,\Delta_i=4}(\delta,\gamma)\;,\label{eq:two_channels_gg}
    \end{equation}
    where the Mellin space conformal blocks in both channels are presented in \cref{sec:conformal_blocks_mellin}, and the $\lambda_{12\mathcal{O}}$ are products of OPE coefficients in either channel. Note there are also similar shifts in the conformal dimension because of superconformal symmetry. As in the 't Hooft loop case, we also record the supergravity amplitude in this formalism, which was found in~\cite{Chen:2026ium}:
    \begin{equation}
        \mathcal{M}_{\text{GG,SUGRA}}(\delta,\gamma) = \frac{2}{(\gamma-1)(\delta-1)}\;.\label{eq:GG_sugra}
    \end{equation}

    \section{Form factors in two languages}\label{sec:approaches}
    \subsection{Stringy corrections from exchanges in Mellin space}\label{sec:mellin_exchanges}
    We are now in a position to compute the stringy corrections to AdS form factors. As we alluded to in the introduction, this is done by matching exchange diagrams on the CFT side to a worldsheet formulation for the form factors in AdS. We will postpone discussion of the worldsheet to \cref{sec:worldsheet}, and focus first on the exchange diagrams.

    We can decompose the form factor, as any other Mellin amplitude, into a sum of exchange diagrams, as in \cref{eq:two_channels_line,eq:two_channels_gg}, which essentially follows from the conformal block decomposition. While additional crossing symmetry, present in defect-free four-point functions, is absent from the defect problem at hand, this decomposition into two different channels turns out to be sufficient and will be essential to our matching problem later.

    We want to obtain corrections to the flat-space limit of the AdS form factors. Thus, we must understand how to take the flat-space limit of these Mellin amplitudes. In the defect-free case, there is a well-known formula, conjectured in~\cite{Penedones:2010ue} and proven in~\cite{Fitzpatrick:2011hu}, which relates Mellin amplitudes to the bulk S-matrix in the flat-space limit of AdS. In the defect case we consider here, a similar formula was first proposed in~\cite{Alday:2024srr} and proven in~\cite{Chen:2025cod}, which relates the Mellin amplitude to the flat-space limit of AdS form factors:
    \begin{equation}
        \mathcal{M}(\delta,\gamma) = \frac{R^{p-d+2}}{\Gamma\left(\frac{\Delta_1+\Delta_2-p}{2}\right)}\int_0^\infty \dd\beta\, \beta^{\frac{1}{2}(\Delta_1+\Delta_2-p)-1}e^{-\beta}\mathcal{A}\left(S = -\frac{2\delta\beta}{R^2},Q=\frac{2\gamma\beta}{R^2}\right),\quad R\to\infty\;,\label{eq:flat_space_general}
    \end{equation}
    where $R$ is the radius of AdS. By the standard AdS/CFT dictionary for $\mathcal{N}=4$ SYM, it is related to the 't Hooft coupling $\lambda$ by
    \begin{equation}
        \frac{\alpha'}{R^2} = \frac{1}{\sqrt{\lambda}}\;.
    \end{equation}
    To consider curvature corrections, we will take \cref{eq:flat_space_general} as the working definition of the form factors in AdS, where we expand in powers of $\frac{1}{R}$ instead of taking the strict $R\to\infty$ limit. For the remainder of this paper, we shall set $\alpha'=1$ and work with $R$ instead of the 't Hooft coupling. Before we can expand the Mellin amplitudes, we should understand how the CFT data, namely the spectrum and OPE coefficients, scale under this limit. 

    Let us begin with the spectrum. As is standard in AdS/CFT~\cite{Witten:1998qj}, the mass of a particle in AdS is related to the conformal dimension and spin on the CFT side by
    \begin{equation}
        (\Delta+\ell)(\Delta+\ell-d)=m^2R^2\label{eq:ads_delta_m}
    \end{equation}
    We will need to consider exchanges which are massive in the $R\to\infty$ limit in order to match to the poles of the worldsheet ansatz. Therefore, $\Delta$ must grow linearly with $R$, i.e. we can write:
    \begin{equation}
        \Delta_{n,\ell} = R\Delta^{(0)}_{n,\ell} + \Delta^{(\frac{1}{2})}_{n,\ell} + \frac{1}{R}\Delta^{(1)}_{n,\ell}+\ldots\;,\label{eq:Delta_R_series}
    \end{equation}
    where $n$ labels states with different conformal dimensions, and this decomposition applies both in the bulk and defect channels. For the OPE coefficients, since the $R\to\infty$ limit of the Mellin amplitudes must scale as $R^0$ to be compatible with flat space, we will choose $\mathcal{N}_{\text{def/bulk},\Delta,\ell}$ in the exchange diagram such that both the exchange diagrams and the OPE coefficients have this same leading scaling, i.e.:
    \begin{equation}
        \lambda_{n,\ell} = \lambda_{n,\ell}^{(0)} + \frac{1}{R}\lambda_{n,\ell}^{(\frac{1}{2})} + \frac{1}{R^2}\lambda_{n,\ell}^{(1)}+\ldots\;,\label{eq:OPE_R_series}
    \end{equation}
    We note that, in principle, there can be a large degeneracy of states with the same conformal dimensions. Our approach here will not be able to distinguish between these different states, and thus we will, in practice, compute averages over them. Thus, the labels $n,\ell$ are all we can use to differentiate different states, and we will denote averages over all other quantum numbers by $\expval{\cdot}$. 
    
    In order to compute the flat-space limit of Mellin amplitudes, we want to consider only quantities which scale as $R^0$. Thus, we define:
    \begin{equation}
        \Delta = R\widehat\Delta,\quad \delta = R^2\hat\delta,\quad \gamma = R^2\hat\gamma\;.
    \end{equation}
    The scaling of $\delta,\gamma$ follows from \cref{eq:flat_space_general}, and $\widehat\Delta$ inherits an expansion at large $R$ from \cref{eq:Delta_R_series}:
    \begin{equation}
        \widehat\Delta_{n,\ell} = \widehat\Delta^{(0)}_{n,\ell} + \frac{1}{R}\widehat\Delta^{(\frac{1}{2})}_{n,\ell} + \frac{1}{R^2}\widehat\Delta^{(1)}_{n,\ell}+\ldots\;,\label{eq:Delta_hat_R_series}
    \end{equation}
    where $\widehat{\Delta}^{(i)}_{n,\ell}=\Delta^{(i)}_{n,\ell}$. In \cref{sec:flat_blocks}, we show how to expand the Mellin representation of the conformal blocks near the flat-space limit. The value of $\widehat{\Delta}_{n,l}^{(0)}$ fixes the location of the poles of the flat-space limit of the conformal blocks. Therefore, in order to match the location of those poles to the flat space result, i.e. to the Veneziano-like amplitude \cref{eq:flat_space_form_factor}, we find that we must set:
    \begin{equation}
        \widehat{\Delta}^{(0)}_{n,\ell,\text{bulk}} = 2\sqrt{n}\;,\quad \widehat{\Delta}^{(0)}_{n,\ell,\text{def}} = \sqrt{n}\;,\label{eq:flat_spectrum}
    \end{equation}
    which is true for both the line defect and giant graviton systems. We expect the form factors to have an expansion in $\frac{1}{R^2}$, rather than $\frac{1}{R}$, and thus demand that the flat-space limit of individual conformal blocks should also have this property. This sets to zero all half-integer terms in the OPE coefficients' expansion \cref{eq:OPE_R_series}, and all but the first half-integer terms in \cref{eq:Delta_R_series,eq:Delta_hat_R_series}. It also completely fixes $\widehat\Delta^{(\frac{1}{2})}_{n,\ell}$:
    \begin{equation}
        \widehat{\Delta}^{(\frac{1}{2})}_{n,\ell,\text{bulk,line}} = \widehat{\Delta}^{(\frac{1}{2})}_{n,\ell,\text{bulk,GG}} = \widehat{\Delta}^{(\frac{1}{2})}_{n,\ell,\text{def,GG}} = -2,\quad \widehat{\Delta}^{(\frac{1}{2})}_{n,\ell,\text{def,line}} = -\frac{3}{2}\;.\label{eq:delta_half}
    \end{equation}
    These are our first result for the curvature corrections to the CFT data. Finally, we can compute the flat-space limit of the SUGRA amplitudes, starting with the line defect case \cref{eq:line_sugra}:
    \begin{equation}
        \begin{aligned}
        \mathcal{A}_{\text{line,SUGRA}}(S,Q;\sigma) ={}& -\frac{8(S+Q\sigma)^2}{SQ}\\
        &+ \frac{8(2QS^2-2S^3-4Q^2S\sigma-4QS^2\sigma+Q^3\sigma^2+2Q^2S\sigma^2)}{Q^2S^2R^2}+\ldots\;,
        \end{aligned}
    \end{equation}
    which has the reduced form:
    \begin{equation}
        \widetilde{\mathcal{A}}_{\text{line,SUGRA}}(S,Q)=\frac{16}{Q} - \frac{32}{Q^2R^2}+\ldots\;.
    \end{equation}
    The fact that this only depends on $Q$ is consistent with the fact that \cref{eq:line_sugra_reduced} only depends on $\gamma$. This amplitude contains the massless ($Q=0$) contributions to the AdS form factors. We chose the normalization here to match with the one we will use for the flat space amplitude in \cref{sec:results}. Similarly, we can compute the SUGRA amplitude for the giant gravitons \cref{eq:GG_sugra} in the flat-space limit:
    \begin{equation}
        \mathcal{A}_{\text{GG,SUGRA}}(S,Q)=-\frac{2}{QS} + \frac{4(Q-S)}{Q^2S^2R^2}+\ldots\;.
    \end{equation}

    \subsection{A worldsheet for AdS form factors}\label{sec:worldsheet}
    In AdS, following~\cite{Alday:2023jdk,Alday:2023mvu}, we assume the existence of a string worldsheet representation for AdS form factors. In general, we assume that this takes the form of an open string worldsheet integral:
    \begin{equation}
        \mathcal{A}(S,Q) = \int_0^1\dd x\, x^{-\frac{S}{2}-1}(1-x)^{-Q-1}\mathcal{G}(S,Q;x)\;,
    \end{equation}
    where $\mathcal{G}(S,Q;x)$ is the worldsheet integrand. Near flat space, we assume that the form factor has an expansion in $\frac{1}{R^2}$, which leads to a similar expansion for the integrand:
    \begin{equation}
        \begin{gathered}
            \mathcal{A}(S,Q) = \mathcal{A}^{(0)}(S,Q) + \frac{1}{R^2}\mathcal{A}^{(1)}(S,Q)+\ldots\;,\\
            \mathcal{G}(S,Q;x) = \mathcal{G}^{(0)}(S,Q;x) + \frac{1}{R^2}\mathcal{G}^{(1)}(S,Q;x)+\ldots
        \end{gathered}
    \end{equation}
    In the strict flat-space limit, the form factors are known to have a worldsheet integral solution~\cite{Hashimoto:1996bf}. The form factor for the line defect in $\mathcal{N}=4$ SYM is given by%
    \footnote{Written as in \cref{eq:flat_space_form_factor}, this corresponds to $a_1=-4Q(t_1\cdot t_2)^2$, $a_2=8S(t_1\cdot\theta)^2(t_2\cdot\theta)^2+16Q(t_1\cdot\theta)(t_2\cdot\theta)(t_1\cdot t_2)$, after pulling out appropriate overall R-symmetry factors, which is the correct flat-space limit of the general form of $a_1,a_2$~\cite{Hashimoto:1996bf,Alday:2024srr}. Our conventions for the normalization of the form factor differ from those of~\cite{Alday:2024srr} by a factor of 2.}
    \begin{equation}
        \mathcal{A}^{(0)}_{\text{line}}(S,Q;\sigma)=-4(S+Q\sigma)^2\frac{\Gamma(-Q)\Gamma\left(-\frac{S}{2}\right)}{\Gamma\left(1-Q-\frac{S}{2}\right)}\;,\label{eq:line_veneziano}
    \end{equation}
    which, in the reduced formalism, becomes
    \begin{equation}
        \widetilde{\mathcal{A}}^{(0)}_{\text{line}}(S,Q) = 8S\frac{\Gamma(-Q)\Gamma\left(-\frac{S}{2}\right)}{\Gamma\left(1-Q-\frac{S}{2}\right)}\;.\label{eq:line_veneziano_reduced}
    \end{equation}
    For the giant gravitons, the form factor only exists in the reduced formalism, and is given by
    \begin{equation}
        \mathcal{A}^{(0)}_{\text{GG}}(S,Q) = -\frac{\Gamma(-Q)\Gamma(-\frac{S}{2})}{\Gamma\left(1-Q-\frac{S}{2}\right)}\;.\label{eq:GG_veneziano}
    \end{equation}
    We can reproduce these with the following worldsheet integrands:
    \begin{equation}
        \widetilde{\mathcal{G}}^{(0)}_{\text{line}}(S,Q;x)=-\frac{\left(64 S-(S-(2Q+S)x)\beta\right)}{4(S+2Q)}\;,\quad \mathcal{G}^{(0)}_{\text{GG}}(S,Q;x) = \frac{2}{S+2Q}\;.\label{eq:flat_integrand}
    \end{equation}
    Here, $\beta$ is an ambiguity parameter. The quantity that it multiplies integrates to exactly 0, and a choice of a value of $\beta$ is equivalent to using the gamma function identities obeyed by the form factor. In deriving the form of these integrands, we assumed that they have an explicit $\frac{1}{S+2Q}$ pole, and that the numerators should be polynomials of homogeneous degree in $S,Q$. This is the form of the integrand on the worldsheet in~\cite{Alday:2023jdk,Alday:2023mvu,Alday:2024yax,Alday:2024ksp}. However, we allow for more general $x$ dependence than those works. This is because of the general form of the worldsheet integral in \cref{eq:flat_space_form_factor}, which has terms which differ by linear functions of $x$.

    Curvature corrections to this worldsheet integrand are built out of special functions known as multiple polylogarithms (MPLs)~\cite{Alday:2023jdk,Alday:2023mvu}. These functions, denoted $L_{w}(x)$, are indexed by words $w$ formed of letters $0,1$. We summarize the construction of MPLs in \cref{sec:mpls}. The length of a word, $|w|$, is known as the transcendentality weight of an MPL. In the first curvature correction, we allow MPLs up to weight 3. This is because $L_{000}(x)=\frac{1}{6}\log^3x$ and $L_{111}(x)=\frac{1}{6}\log^3(1-x)$, and the flat-space limit of conformal blocks have at most quartic poles (see \cref{sec:flat_blocks}). We show in \cref{sec:mpls} that $\log^bx$ integrates to a pole of order $b+1$ in $S$, and similarly with $\log^b(1-x)$ in $Q$. Thus, this is the minimal basis of MPLs that reproduces all poles that can appear in the form factor. 
    
    Finally, we note that $\zeta(b)L_w(x)$ should be understood as an MPL of transcendentality weight $b+|w|$. This is because the MPLs are directly linked to the values of the $\zeta$ function, which are the values taken by $L_w(1)$. Thus, the full basis of MPLs also includes $\zeta$ functions multiplying lower-weight MPLs. It is convenient to organize a basis of MPLs as vectors indexed by the weight:
    \begin{equation}
        \begin{aligned}
        \vec{L}_3(x)=\Big(&L_{000}(x),L_{001}(x),L_{010}(x),L_{011}(x),L_{100}(x),L_{101}(x),\\
        &L_{110}(x),L_{111}(x),\zeta(2)L_0(x),\zeta(2)L_1(x),\zeta(3)\Big)\;,\\
        \vec{L}_2(x) = \Big(&L_{00}(x),L_{01}(x),L_{10}(x),L_{11}(x),\zeta(2)\Big)\;,\\
        \vec{L}_1(x) = \Big(&L_0(x),L_1(x)\Big)\;,\\
        \vec{L}_0(x)=\Big(&1\Big)\;.
        \end{aligned}
    \end{equation}
    In the case of the AdS Veneziano amplitude, there is a further restriction on the ansatz. MPLs, as generalizations of logarithms, possess branch cuts. However, there are related functions, known as single-valued MPLs (SVMPLs), denoted $\mathcal{L}_w(z)$ which are defined on the complex plane and built of linear combinations of MPLs evaluated at $z$ and $\bar z$ such that the branch cuts cancel. The simplest example is $\mathcal{L}_0(z)=\log(z\bar z)$. While open string worldsheet integrands are not single-valued, and thus are not built of SVMPLs, they can be built out of the ``diagonal limit'' of SVMPLs~\cite{Alday:2024yax,Alday:2024ksp}, which is to say SVMPLs where $z=\bar z=x$, which we denote $\mathcal{L}_w^{\text{diag}}(x)$. We also consider a second, more restrictive, basis of diagonal limits of SVMPLs%
    \footnote{Not all words that could naively show up at each weight appears in this basis. This is because, in the diagonal limit, SVMPLs with a word $w$ and the reverse of that word are equivalent, see \cref{sec:mpls}.}:
    \begin{equation}
        \begin{aligned}
            \vec{\mathcal{L}}_3(x) = \Big(&\mathcal{L}_{000}^{\text{diag}}(x),\mathcal{L}_{001}^{\text{diag}}(x),\mathcal{L}_{010}^{\text{diag}}(x),\mathcal{L}_{011}^{\text{diag}}(x),\mathcal{L}_{101}^{\text{diag}}(x),\mathcal{L}_{111}^{\text{diag}}(x),\\
            &\zeta(2)\mathcal{L}_0^{\text{diag}}(x),\zeta(2)\mathcal{L}_1^{\text{diag}}(x),\zeta(3)\Big)\;,\\
            \vec{\mathcal{L}}_2(x) = \Big(&\mathcal{L}_{00}^{\text{diag}}(x),\mathcal{L}_{01}^{\text{diag}}(x),\mathcal{L}_{11}^{\text{diag}}(x),\zeta(2)\Big)\;,\\
            \vec{\mathcal{L}}_1(x)=\Big(&\mathcal{L}_0^{\text{diag}}(x),\mathcal{L}_1^{\text{diag}}(x)\Big)\;,\\
            \vec{\mathcal{L}}_0(x)=\Big(&1\Big)\;.\label{eq:mpl_basis}
        \end{aligned}
    \end{equation}
    Then, we define vectors $P_j^a(S,Q;x)$ and $\vec{\mathcal{P}}_j^a(S,Q;x)$ , which have the same length as the $\vec{L}_j^a(x)$ and $\vec{\mathcal{L}}_j(x)$, respectively. Each component of these vectors is a polynomial of homogeneous degree $a$ in $S,Q$, and linear in $x$, with rational coefficients. In terms of these vectors, our ansatz for the worldsheet integrand is then:
    \begin{equation}
        \widetilde{\mathcal{G}}^{(1)}_{\text{line}}(S,Q;x) = \sum_{j=0}^3 \frac{\vec{\mathcal{P}}_j^{j+1}(S,Q;x)\cdot \vec{\mathcal{L}}_j(x)}{(S+2Q)^2}\;,\quad \mathcal{G}^{(1)}_{\text{GG}}(S,Q;x) = \sum_{j=0}^3 \frac{\vec{P}_j^{j}(S,Q;x)\cdot \vec{L}_j(x)}{(S+2Q)^2}\;.\label{eq:integrand_ansatz}
    \end{equation}
    For the line defect, we use the more restrictive ansatz built from diagonal limits of SVMPLs. This ansatz is sufficient to uniquely fix the form factor, and the resulting answer is compatible with the independent supersymmetric localization constraints discussed in the next section. For the giant graviton, however, we find that the corresponding restricted ansatz is incompatible with the localization result. We therefore enlarge the ansatz to general MPLs. With this larger ansatz, pole matching leaves five nontrivial coefficients unfixed, as we discuss in \cref{subsec:giantgraviton}.

    \section{Results}\label{sec:results}
    We finally present the details of our matching between the poles coming from the flat-space limit of the CFT, and the worldsheet. Our procedure for the comparison is to expand both sides near each pole, and compare order-by-order. The poles at $S,Q=0$ are given by the SUGRA amplitudes \cref{eq:line_sugra_reduced,eq:GG_sugra}. The comparison is always performed in the reduced correlators.

    Exchanged operators are organized by their Regge trajectory, which is encoded by the relationship between $n$ and $\ell$. In the bulk channel, the $j$-th Regge trajectory obeys $\ell=2(n-j)$, while in the defect channel it obeys $\ell=n-j$. We will present our CFT data by Regge trajectory. In both the line defect and giant graviton cases, the first Regge trajectory in the bulk channel is given by the Konishi-like operators, which have known conformal dimension~\cite{Gromov:2011de,Basso:2011rs,Gromov:2011bz}:
    \begin{equation}
        \widehat{\Delta}_{n,2(n-1),\text{bulk}} = 2\sqrt{n} - \frac{2}{R} + \frac{3n^2-n+2}{2\sqrt{n}R^2}+\ldots\;,\label{eq:konishi}
    \end{equation}
    which we note, in particular, obeys \cref{eq:delta_half,eq:flat_spectrum}. In the line defect case, the worldsheet matching is sufficient to recover this result. In the giant graviton case, however, we use it as an input. The Konishi-like trajectory is known to be non-degenerate. For other trajectories, however, the operators may have degeneracy (in the bulk channel, it is known that they do), and thus we can only obtain the sum of any CFT data over all degenerate operators, which we denote by:
    \begin{equation}
        \expval{f_{n,\ell}} = \sum_r f_{n,\ell,r}\;,
    \end{equation}
    where $r$ are any other quantum numbers labelling degenerate states, and $f$ is any type of CFT data.

    \subsection{Line defect}
    We begin by presenting the solution we find for the worldsheet integral after matching massive poles (from the flat-space limit of conformal blocks as in \cref{sec:flat_blocks}) and massless poles (from SUGRA as in \cref{eq:line_sugra_reduced}). In terms of \cref{eq:integrand_ansatz}, we find:
    \begin{equation}
        \begin{aligned}
        \vec{\mathcal{P}}_3^4(S,Q;x)=\Big(&-\frac{1}{2}S^3(S+2Q),2S^2(S^2+3QS+2Q^2),S^2(2S^2+9SQ+10Q^2),\\
        &8SQ^2(S+2Q),4SQ(S+2Q)^2,-8SQ^2(S+2Q),0,0,-4S^3(S+2Q)\Big)\;,\\
        \vec{\mathcal{P}}_2^3(S,Q;x)=\Big(&-5S^2(S+2Q),4S(5S^2-2SQ-20Q^2),16SQ(5S+9Q),0\Big)\;,\\
        \vec{\mathcal{P}}_1^2(S,Q;x)=\Big(&8(10S^2+25SQ+14Q^2)x,-16(7S^2+6SQx+14Q^2x)\Big)\;,\\
        \vec{\mathcal{P}}_0^1(S,Q;x)=\Big(&-32(4S-Q(26x+7))\Big)\;.
        \end{aligned}\label{eq:line_worldsheet_solution}
    \end{equation}
    On top of these solutions, there are 53 linear combinations allowed by our ansatz which integrate to 0. These should be understood to parametrize generalizations of the gamma function identities allowed by the flat-space form factor, and are thus generalizations of the $\beta$ parameter in \cref{eq:flat_integrand}. These are the equivalent of the ambiguities that show up in the AdS Virasoro-Shapiro worldsheet integrand~\cite{Alday:2023mvu}. Thus, we could add any multiple of these to the answers above without modifying the final form factor. We chose values for these added ambiguities that make the answer above as simple as we could.

    Once this integrand is established, we can, in principle, extract an infinite amount of CFT data from it, in the form of the flat space and first curvature corrections for the OPE coefficients and spectrum. We get the $\lambda_{n,\ell}^{(0)}$ from the flat-space integral \cref{eq:flat_integrand}, and the $\lambda_{n,\ell}^{(1)}$ and $\Delta_{n,\ell}^{(1)}$ from the solution for $\widetilde{\mathcal{G}}^{(1)}_{\text{line}}(S,Q)$. The first data we can obtain are the flat space OPE coefficients:
    \begin{equation}
        \begin{gathered}
            \expval{\lambda^{(0)}_{n,2(n-1),\text{bulk,line}}} = \frac{2^{n+4}n^{n/2+1}\sqrt{\pi}}{\Gamma(n)}\;,\quad \expval{\lambda^{(0)}_{n,2(n-3),\text{bulk,line}}} = \frac{2^{n+4}n^{n/2+1}\sqrt{\pi}(n-2)(n-1)(n+3)}{96\Gamma(n)(2n-3)}\;,\\
            \expval{\lambda^{(0)}_{n,n-1,\text{def,line}}} = \frac{2^{8-3q-n}n^n}{\sqrt{\pi}n^{1-q}(2n-3)!!}\;,\quad \expval{\lambda^{(0)}_{n,n-3,\text{def,line}}} = \frac{2^{8-3q-n}n^n(2n-5)(n+3)}{6\sqrt{\pi}n^{2-q}(2n-3)!!}\;,
        \end{gathered}
    \end{equation}
    where $q$ is the parity of $n$, i.e.:
    \begin{equation}
        n \equiv q \mod 2\;.
    \end{equation}
    We can also obtain $\frac{1}{R^2}$ corrections to the OPE data. For example, in the defect channel, we find the following data for the spectrum of the leading Regge trajectory:
    \begin{equation}
        \frac{\expval{\lambda^{(0)}_{n,n-1,\text{def,line}}\widehat{\Delta}_{n,n-1,\text{def,line}}^{(1)}}}{\expval{\lambda^{(0)}_{n,n-1,\text{def,line}}}} = \frac{6n^2-2n+9}{8\sqrt{n}}\;.
    \end{equation}
    If the leading Regge trajectory in the bulk channel is non-degenerate, as is the case for the Konishi-like Regge trajectory in the bulk channel, the OPE coefficient cancels and this is just the anomalous dimension of that Regge trajectory. As stated above, we find that the bulk channel's first Regge trajectory's anomalous dimensions are given by \cref{eq:konishi}. We can also fix similar data for the third Regge trajectories in both channels:
    \begin{equation}
        \frac{\expval{\lambda^{(0)}_{n,2(n-3),\text{bulk,line}}\widehat{\Delta}_{n,2(n-3),\text{bulk,line}}^{(1)}}}{\expval{\lambda^{(0)}_{n,2(n-3),\text{bulk,line}}}} = \frac{n(9n+1)}{6\sqrt{n}}\;,\quad \frac{\expval{\lambda^{(0)}_{n,n-3,\text{def,line}}\widehat{\Delta}_{n,n-3,\text{def,line}}^{(1)}}}{\expval{\lambda^{(0)}_{n,n-3,\text{def,line}}}} = \frac{18n^2+2n+15}{24\sqrt{n}}\;,
    \end{equation}
    where degeneracy is expected. Finally, we obtain corrections to the OPE coefficients:
    \begin{equation}
        \begin{gathered}
        \expval{\lambda^{(1)}_{n,2(n-1),\text{bulk,line}}} = \frac{2^{n+4}n^{n/2+1}\sqrt{\pi}}{\Gamma(n)}\left(\frac{3n^4-46n^3+105 n^2-158 n + 24}{24 n} + n^2\zeta(3)\right)\;.\\
        \expval{\lambda_{n,n-1,\text{def,line}}^{(1)}} = \frac{2^{8-3q-n}n^n}{\sqrt{\pi}n^{1-q}(2n-3)!!}\left(\frac{-7 n^3+6 n^2-3 (-1)^n \left(3 n^2-n+3\right)-23n-15}{12 n}+n^2\zeta(3)\right)\;.
        \end{gathered}
    \end{equation}
    We only present data from odd Regge trajectories because the flat space OPE coefficients for the even ones always vanish. Indeed, we get:
    \begin{equation}
       \expval{\lambda_{n,2(n-m),\text{bulk,line}}^{(0)}}=\expval{\lambda_{n,n-m,\text{def,line}}^{(0)}}=0\;,\quad m\ \text{even}\;.
    \end{equation}
    In the defect channel, this remains true at the leading curvature correction:
    \begin{equation}
        \expval{\lambda_{n,n-m,\text{def,line}}^{(1)}}=0\;,\quad m\ \text{even}\;.
    \end{equation}
    However, in the bulk channel, we find that the OPE coefficients are now non-zero, and, in fact, negative%
    \footnote{Note that, in the bulk channel, the OPE coefficients that appear are not squares of real numbers, so their sign is not determined by unitarity}:
    \begin{equation}
        \expval{\lambda_{n,2(n-m),\text{bulk,line}}^{(1)}}<0\;,\quad m\ \text{even}\;.
    \end{equation}
    For example, the first such trajectories have:
    \begin{equation}
        \expval{\lambda^{(1)}_{n,2(n-2),\text{bulk,line}}} = -5\frac{2^n n^{(n+1)/2}\sqrt{\pi}}{\Gamma(n-1)}\;,\quad \expval{\lambda^{(1)}_{n,2(n-4),\text{bulk,line}}} = -5\frac{2^n n^{(n+1)/2}(n-3)(n-2)(n+5)\sqrt{\pi}}{96\Gamma(n-1)(2n-5)}\;.
    \end{equation}
    This means that new operators appear in the exchange at $\frac{1}{R^2}$. We can make no predictions about the spectrum of these operators, as the anomalous dimensions can only appear at order $\frac{1}{R^4}$.

    \subsubsection{Supersymmetric localization check}\label{sec:line_localization}
    We now want to check that our results are consistent with what is known in the literature. In particular, there are two quantities, denoted by $\mathcal{I}$ and $\mathcal{E}$, which can be computed exactly by supersymmetric localization as functions of the complex coupling $\tau=\frac{\theta}{2\pi}+\frac{4\pi i}{g^2}=\tau_1+i\tau_2$~\cite{Pufu:2023vwo,Billo:2023ncz,Dempsey:2024vkf,Billo:2024kri}. If we denote the line defect by $\mathbb{L}$, the first of these is the second derivative of its expectation value with respect to the mass, while the second is given by the derivatives with respect to the coupling:
    \begin{equation}
        \mathcal{I}(\tau,\bar\tau) = \eval{\partial^2_m\log\expval{\mathbb{L}}}_{m=0}\;,\quad \mathcal{E}(\tau,\bar\tau) = \eval{\partial_\tau\partial_{\bar\tau}\log\expval{\mathbb{L}}}_{m=0}\;.
    \end{equation}
    The first is related to an integral of the $\sigma^2$ coefficient of the correlator, $\mathcal{F}_2(\zeta,\eta)$~\cite{Dempsey:2024vkf}:
    \begin{equation}
        \mathcal{I} = -16\pi^4\int_1^\infty\dd\zeta\int_{-1}^1\dd\eta\, \frac{N^2}{4\pi^4}\mathcal{F}_2(\zeta,\eta)\;.
    \end{equation}
    The second one is related to the topologically twisted correlator:
    \begin{equation}
        \mathcal{E}(\tau,\bar\tau) = \frac{c}{8\tau_2^2}\left(\mathcal{F}_{\text{top}}(-1)-a(\tau,\bar\tau)^2\right)\;,
    \end{equation}
    where $\mathcal{F}_{\text{top}}=\mathcal{F}(z,z;z)$ depends only on the sign of $z$, as discussed in \cref{sec:line_defect_setup}. Here, $c=\frac{N^2-1}{4}$ and $a(\tau,\bar\tau)$ is the one point function coefficient:
    \begin{equation}
        \dexpval{\mathcal{O}_2(x,t)} = a(\tau,\bar\tau)\frac{(t\cdot\theta)^2}{|x^i|^2}\;.
    \end{equation}
    This can also be computed by localization:
    \begin{equation}
        a(\tau,\bar\tau) = \eval{-i\frac{\tau_2}{\sqrt{2c}}\partial_\tau\log\expval{\mathbb{L}}}_{m=0}\;.
    \end{equation}
    This allows us to interpret $\mathcal{E}(\tau,\bar\tau)$ as the connected part of the twisted correlator where we extract the identity operator in the defect channel:
    \begin{equation}
        \mathcal{E}(\tau,\bar\tau) = \frac{2c}{\tau_2^2}\mathcal{F}_{\text{top,conn}}(-1)\;.
    \end{equation}
    In our case, the integrated correlator $\mathcal{I}(\tau,\bar\tau)$ was computed in~\cite{Pufu:2023vwo}:
    \begin{equation}
        \mathcal{I}(\tau,\bar\tau)=N\left(\frac{4\pi}{\lambda^{\frac{1}{2}}}-\frac{2\pi^3}{\lambda^{\frac{3}{2}}}+\frac{24\pi\zeta(3)}{\lambda^2}+\frac{\pi^5}{6\lambda^{\frac{5}{2}}}+O(\lambda^{-3})\right)+O(N^0)\;.
    \end{equation}
    For $\mathcal{E}(\tau,\bar\tau)$, the relevant result was computed in~\cite{Alday:2024srr}, using a result of~\cite{Drukker:2000rr} for the vacuum expectation value of the 't Hooft loop:
    \begin{equation}
        \begin{aligned}
            \mathcal{E}(\tau,\bar\tau) ={}& \frac{1}{N}\frac{\lambda(3\lambda+4\pi^2)}{64\pi\sqrt{\lambda+\pi^2}}+O(N^{-2})\\
            ={}&\frac{1}{N}\left(\frac{3\lambda^{\frac{3}{2}}}{64\pi}+\frac{5\pi\lambda^{\frac{1}{2}}}{128}-\frac{7\pi^3}{512\lambda^{\frac{1}{2}}}+\frac{9\pi^5}{1024\lambda^{\frac{3}{2}}}+O(\lambda^{-\frac{5}{2}})\right)+O(N^{-2})\;.
        \end{aligned}
    \end{equation}
    In order to use these results to obtain checks on our worldsheet integrand, we can consider the stringy AdS form factor in a low energy expansion. It was shown in \cite{Alday:2024srr} that the stringy corrections to the 't Hooft loop two-point function can be organized as 
    \begin{equation}\label{linedefectlowenergyexp}
        \mathcal{F}(z,\bar z;\sigma) = \mathcal{F}_{\text{sugra}}(z,\bar z;\sigma)f(\lambda) + \lambda^{-1}\mathcal{F}^{\text{h.d.}}_{L=2}(z,\bar z;\sigma)+\lambda^{-\frac{3}{2}}\mathcal{F}^{\text{h.d.}}_{L=3}(z,\bar z;\sigma)+\lambda^{-2}\mathcal{F}^{\text{h.d.}}_{L=4}(z,\bar z;\sigma)+\ldots\;,
    \end{equation}
    which has the form of a supergravity solution with a $\lambda$ dependent coefficient plus an infinite tower of higher-derivative corrections. Here $L$ labels the derivative cutoff and the higher-derivative correction term $\mathcal{F}^{\text{h.d.}}_L$ contains defect contact Witten diagrams with at most $2L$ derivatives. The supergravity solution is multiplied by a coupling dependent factor ~\cite{Alday:2024srr}:
    \begin{equation}
        f(\lambda) = \frac{a(\lambda)}{a(\lambda)|_{\text{leading}}} = \sqrt{1+\frac{\pi^2}{\lambda}}\;.
    \end{equation}
This happens because the defect one-point function is not protected. The $1/\lambda$ expansion of the defect correlator can be translated into Mellin space:
    \begin{equation}
        \mathcal{M}(\delta,\gamma;\sigma) = \mathcal{M}_{\text{sugra}}(\delta,\gamma;\sigma)f(\lambda) + \lambda^{-1}\mathcal{M}^{\text{h.d.}}_{L=2}(\delta,\gamma;\sigma)+\lambda^{-\frac{3}{2}}\mathcal{M}^{\text{h.d.}}_{L=3}(\delta,\gamma;\sigma)+\lambda^{-2}\mathcal{M}^{\text{h.d.}}_{L=4}(\delta,\gamma;\sigma)+\ldots\label{eq:higher_derivatives_mellin}
    \end{equation}
    In particular, the higher-derivative correction terms become polynomials of degree $L$ and the SCWI restrict the possible polynomials that can appear. At the first cutoff order with $L=1$, there are no polynomial solutions to the SCWI, which explains the absence of such a term above. At the second order with $L=2$, there is only one possible solution up to rescalings, while there are two additional solutions of order three and 3 new solutions at order four. We collect our basis for these solutions, as well as their representation as reduced Mellin amplitudes, in \cref{sec:scwi_mellin}. In the basis defined in that appendix, we define:
    \begin{equation}
        \begin{gathered}
            \mathcal{M}_{L=2}^{\text{h.d.}} = \frac{\mathcal{N}_{\text{sugra}}}{\pi}c_1^{(2)}M^{(1)}_{L=2}\;,\\
            \mathcal{M}_{L=3}^{\text{h.d.}} = \frac{\mathcal{N}_{\text{sugra}}}{\pi}\left(c_1^{(3)}M_{L=3}^{(1)}+c_2^{(3)}M_{L=3}^{(2)}+c_3^{(3)}M_{L=2}^{(1)}\right)\;,\\
            \mathcal{M}_{L=4}^{\text{h.d.}} = \frac{\mathcal{N}_{\text{sugra}}}{\pi}\left(c^{(4)}_1M^{(1)}_{L=4}+c^{(4)}_2M^{(2)}_{L=4}+c^{(4)}_3M^{(3)}_{L=4}+c^{(4)}_4M^{(1)}_{L=3}+c^{(4)}_5M^{(2)}_{L=3}+c^{(4)}_6M^{(1)}_{L=2}\right)\;,\label{eq:localization_higher_derivs}
        \end{gathered}
    \end{equation}
    where $\mathcal{N}_{\text{sugra}}$ is a normalization factor introduced in~\cite{Alday:2024rjs}. We can then plug these into the flat-space limit formula and compare them to the small-$\alpha'$ expansion of the worldsheet answer \cref{eq:line_veneziano}, or \cref{eq:line_veneziano_reduced} in the reduced formalism. We can also, in parallel, use the position space form of this ansatz to obtain $\mathcal{I}$ and $\mathcal{E}$, and compare to the answers from localization. This analysis was done in~\cite{Alday:2024srr}, of which the results we will quote below. The flat-space limit fixes the coefficients of the leading solutions to be:
    \begin{equation}
        \begin{gathered}
        c_1^{(2)} = \frac{5\pi^2}{4}\;,\quad c_1^{(3)}=\frac{105}{2}\zeta(3)\;,\quad c_2^{(3)}=-\frac{105}{4}\zeta(3)\;,\\
        c_1^{(4)} = \frac{21\pi^4}{4}\;,\quad c_2^{(4)}=-\frac{21\pi^4}{32}\;,\quad c_3^{(4)}=\frac{21\pi^4}{16}\;,
        \end{gathered}
    \end{equation}
    where the first of these could also have been fixed by either of the localization constraints, such that there were three nontrivial conditions for a single number. Localization then further provides the following restrictions:
    \begin{equation}
        c_3^{(3)}=-50\zeta(3)\;,\quad c_5^{(4)}=-\frac{17269\pi^4}{1440}-c_4^{(4)}\;,\quad c_6^{(4)}=-\frac{157\pi^4}{72}-\frac{20}{21}c_4^{(4)}\;,
    \end{equation}
    where we note that the eight derivative terms are fixed up to an undetermined coefficient. Let us compare this with the leading curvature correction computed using the worldsheet ansatz. Expanding the worldsheet solution%
    \footnote{In this equation, we used the supergravity limit to set the normalization to be the same as the one we use for the higher derivative corrections.}
    in the low energy ($\alpha'\to 0$) limit, we get:
    \begin{equation}
        \widetilde{\mathcal{A}}^{(1)}(\alpha'S,\alpha' Q) = \mathcal{N}_{\text{sugra}}\left(\frac{-8}{\pi^{\frac{3}{2}}Q^2(\alpha')^2}+\frac{25S}{\pi^{\frac{3}{2}}}\zeta(3)\alpha' - \frac{S(46S-461Q)\pi^{\frac{5}{2}}}{360}(\alpha')^2+\ldots\right)\;.
    \end{equation}
    On the other hand, the stringy form factor built from localization has the form:
    \begin{equation}
        \begin{aligned}
        \widetilde{\mathcal{A}}_{\text{loc}}(\alpha'S,\alpha'Q) = \mathcal{N}_{\text{sugra}}\Bigg(&\frac{-8}{\pi^{\frac{3}{2}}Q^2(\alpha')^2}+\frac{25S}{\pi^{\frac{3}{2}}}\zeta(3)\alpha'\\
        &- \frac{\left((7504S-10241Q)\pi^4+480c_4^{(4)}(S+Q)\right)S}{12600\pi^{\frac{3}{2}}}(\alpha')^2+\ldots\Bigg)\;.
        \end{aligned}
    \end{equation}
    The SUGRA part, i.e. the power of $(\alpha')^{-2}$, is not a check on our answer, as we have already used SUGRA as an input for our answer. However, the $\alpha'$ piece matching is a first nontrivial check. There is a second check given by the fact that the coefficients of $SQ$ and $S^2$ in the $(\alpha')^2$ piece give two equations for a single number, which has the solution:
    \begin{equation}
        c_4^{(4)}=-\frac{2947}{240}\pi^4\;.
    \end{equation}
    We emphasize that these two equations having a solution provides a nontrivial check on our solution for the form factor.

    \subsection{Giant gravitons}\label{subsec:giantgraviton}
    We now consider the giant graviton system. Unlike for the line defect, the worldsheet bootstrap by itself does not completely fix the form factor. We will first use supersymmetric localization as an additional input, which fixes one of the remaining parameters. We will then show that an alternative physical assumption--that no new operators appear in the defect channel at this order--completely fixes the form factor, in which case localization instead provides an independent check.
        
    In the giant graviton system, there is a single integrated correlator, rather than two as in the line defect case:
    \begin{equation}
        \mathcal{I} = -\frac{2N}{\pi}\int_0^\infty\dd r\int_0^\pi \dd\theta\frac{r^3\sin^2\theta}{U^2}\left(\frac{x_{12}^2x_{13}^2x_{24}^2}{x_{34}^2}\mathcal{H}(r,\theta)\right)\;,
    \end{equation}
    where:
    \begin{equation}
        U=1-2r\cos\theta+r^2\;,\quad V=r^2\;.
    \end{equation}
    The integrated correlator can be computed from localization, and at strong coupling admits the expansion~\cite{Brown:2024tru}%
    \footnote{Our answer differs from~\cite{Brown:2024tru} by a factor of 2, which can be attributed to a difference in normalization.}:
    \begin{equation}
        \mathcal{I} = 1-\frac{2\pi^2}{3\lambda}-\frac{4\zeta(3)}{\lambda^{\frac{3}{2}}}+O(\lambda^{-\frac{5}{2}})\;,\label{eq:localization_higher_derivs_gg}
    \end{equation}
    where we note that all higher integer powers of $\lambda$ vanish. 
    
    As in the previous part, we want to compute stringy corrections to SUGRA in the low energy expansion in order to compare to our worldsheet results. Structurally, this expansion is similar to \cref{linedefectlowenergyexp} but also contains a few important differences which we should highlight. The supergravity term has a fixed coefficient 1 rather than a coupling dependent coefficient, and this is because the defect one-point function is protected for giant gravitons. Moreover, in order to find pure contact diagram solutions to the superconformal Ward identities, we need to include higher-derivative contact Witten diagrams with both even and odd numbers of derivatives below a certain derivative cutoff.\footnote{While contact Witten diagrams with odd number derivatives are necessary for finding superconformal solutions, new solutions only show up when the number of derivatives cutoff is even, i.e., at $2L=4, 6, 8,\ldots$.} However, when rewritten in terms of the reduced Mellin amplitude, these are again just polynomials of the Mellin-Mandelstam variables and the counting of independent solutions is the same as in the line defect case. Using the strict flat-space limit and localization, we can completely fix the low energy expansion up to six derivatives but the corrections with more derivatives are under-constrained. Explicitly, the first few orders of the low energy expansion, obtained using localization as input, is:
    \begin{equation}
        \mathcal{A}^{(1)}_{\text{loc}}(\alpha'S,\alpha'Q)=-\frac{4(S-Q)}{Q^2S^2(\alpha')^3} + \zeta(3) +\ldots\label{eq:A1_from_localization_GG}
    \end{equation}
    Already, at the next order in $\alpha'$, which is linear in $S,Q$, localization does not impose any constraints. Pole matching, together with the assumption that the Konishi trajectory is the leading Regge trajectory in the bulk channel, fixes the form factor up to five coefficients. In addition, there are 45 ambiguities at the level of the worldsheet integrand which integrate to zero and therefore do not affect the form factor. Matching to the localization result in the low-energy limit fixes one of the five physical coefficients, leaving four unfixed coefficients, which we denote by $C_i$. In terms of these coefficients, and in the form of \cref{eq:integrand_ansatz}, the result is:
    \begin{subequations}
        \begin{align}
        \begin{split}
            \vec{P}_3^3(S,Q;x) ={}& \Big(\frac{1}{2} S^2 (2Q+S),\\
            &\frac{1}{10} S (2 Q+S)\left(\left(22 C_1+9 C_2-9C_4-30\right) Q+2 \left(4 C_1+3 C_2-3C_4-5\right) S\right),\\
            &-\frac{1}{20} S(2 Q+S) \left(\left(6 C_1-3 C_2-20C_3+3 C_4+50\right) Q-4 \left(4 C_1+3C_2-3 C_4-5\right) S\right),\\
            &-4 Q^2 (2Q+S),-S \left(2 Q^2+3 QS+S^2\right),\\
            &-\frac{1}{5} Q (2 Q+S)\left(\left(6 C_1-3 C_2+3C_4+20\right) Q+3 \left(2C_1-C_2+C_4\right)S\right),\\
            &-\frac{1}{5} Q (2 Q+S)\left(\left(6 C_1-3 C_2-20 C_3+3C_4+20\right) Q+\left(-2 C_1-9 C_2+10C_3+9 C_4-5\right) S\right),\\
            &8 Q^2 (2Q+S),0,-\frac{1}{5} \left(8 C_1+6C_2-10 C_3-6 C_4+5\right) Q S (2Q+S),\\
            &-\frac{1}{5} \left(4 C_1+3 C_2-3C_4\right) S^2 (2 Q+S)\Big)\;,
        \end{split}\\
        \begin{split}
            \vec{P}_2^2(S,Q;x) = {}& \Big(S (2 Q+S),\\
            &-\frac{1}{5} \big(2\left(18 C_1+16 C_2-6 C_4-65\right)Q^2+\left(26 C_1+42 C_2-12C_4-85\right) Q S\\
            &+\left(4 C_1+13C_2-3 C_4-5\right)S^2\big),\\
            &-\frac{1}{10} \big(-2\left(6 C_1-3 C_2-20 C_3+3C_4+110\right) Q^2+2 \left(26 C_1+42C_2+40 C_3-12 C_4-55\right) Q S\\
            &+5\left(4 C_1+9 C_2-3 C_4+5\right)S^2\big),-8 Q (5 Q+3S),\\
            &\frac{1}{10} S \left(\left(-8C_1+14 C_2+40 C_3+6 C_4+50\right)Q+\left(12 C_1+19 C_2-9 C_4+35\right)S\right)\Big)\;,
        \end{split}\\
        \begin{split}
            \vec{P}_1^1(S,Q;x) ={}&\Big(-\frac{1}{5} \left(Q\left(\left(18 C_1-9 C_2+60 C_3+9C_4-50\right) x+60\right)+\left(4C_1+33 C_2-3 C_4+85\right)S\right),\\
            &-\frac{2}{5} Q \left(10\left(4 C_2+3\right) x+4 C_1-7 C_2-3C_4-55\right)-4 S \left(2 C_2 x-2\left(C_2+3\right)+x\right)\Big)\;,
        \end{split}\\
        \begin{split}
            \vec{P}_0^0(S,Q;x)={}&\Big(\frac{2}{5}\left(4C_1+33 C_2-3C_4+65\right)\Big)\label{eq:gg_worldsheet_solution_full}
        \end{split}
        \end{align}
    \end{subequations}
    We can now present CFT data. The first few flat space OPE coefficients in the bulk channel are given by:
    \begin{equation}
        \begin{gathered}
        \expval{\lambda^{(0)}_{n,2(n-1),\text{bulk,GG}}} = \frac{2^nn^{n+1}}{\Gamma(n)}\;,\quad \expval{\lambda^{(0)}_{n,2(n-2),\text{bulk,GG}}} = \frac{2^nn^{n+1}}{8\Gamma(n)}\;,\\
        \expval{\lambda^{(0)}_{n,2(n-3),\text{bulk,GG}}} = \frac{2^nn^{n}(n-1)(n+4)}{192\Gamma(n)}\;, \expval{\lambda^{(0)}_{n,2(n-4),\text{bulk,GG}}} = \frac{2^nn^{n}(n-1)(n+4)}{1536\Gamma(n)}\;.
        \end{gathered}
    \end{equation}
    Unlike the line defect case, we find that the even Regge trajectories do not vanish. Instead, they are determined by the odd trajectories:
    \begin{equation}
        \expval{\lambda^{(0)}_{n,2(n-m),\text{bulk,GG}}} = \frac{1}{8}\expval{\lambda^{(0)}_{n,2(n-m+1),\text{bulk,GG}}}\;,\quad m\ \text{even}\;.
    \end{equation}
    In the defect channel, the even trajectories vanish once again, and we have the following data for the first trajectories:
    \begin{equation}
        \begin{gathered}
        \expval{\lambda^{(0)}_{n,n-1,\text{def,GG}}} = \frac{9 n^{n-2+q}2^{11-2n-5q}}{\Gamma(n)}\;,\quad \expval{\lambda^{(0)}_{n,n-3,\text{def,GG}}} = \frac{3n^{n-4+q}2^{11-2n-5q}(n+2)(n-2)^2}{\Gamma(n)}\;,\\
        \expval{\lambda^{(0)}_{n,n-m,\text{def,GG}}}=0\;,\quad m\ \text{even}\;.
        \end{gathered}
    \end{equation}
    The $C_i$ were chosen to make the leading anomalous dimensions as simple as possible. For the first Regge trajectories in the defect channel, we have:
    \begin{equation}
        \frac{\expval{\lambda^{(0)}_{n,n-1,\text{def,GG}}\widehat{\Delta}_{n,n-1,\text{def,GG}}^{(1)}}}{\expval{\lambda^{(0)}_{n,n-1,\text{def,GG}}}} = \frac{2(C_1+\zeta(2)C_3)n^2+2C_2n+1}{2\sqrt{n}}\;.
    \end{equation}
    The expressions for higher CFT data are considerably more complicated, as they depend on all of the $C_i$. We present some of these results in \cref{sec:gg_CFT_data}. 
    
    The remaining four-parameter freedom has a simple interpretation in the defect-channel CFT data. For generic values of the $C_i$, new operators appear at order $1/R^2$: in particular, we find nonzero values of $\langle\lambda^{(1)}_{n,n-2,\mathrm{def},GG}\rangle$ for the even Regge trajectories, which were absent at leading order. Unitarity requires the corresponding defect-channel OPE coefficients to be non-negative. There is a distinguished point in coefficient space,
    \begin{equation}\label{solC1234}
        C_1 = \frac{3}{4}\;,\quad C_2 = -\frac{3}{4}\;,\quad C_3 = 0\;,\quad C_4 = \frac{13}{12}\;.
    \end{equation}
    At this point, all $1/R^2$ corrections to the OPE coefficients of the even Regge trajectories vanish, so that no new operators appear in the defect channel.
    
Remarkably, requiring the absence of such new operators at order $1/R^2$, as was the case for the line defect, uniquely fixes all four remaining coefficients. This condition can be imposed without using localization and selects precisely the point \cref{solC1234}. The resulting form factor reproduces the localization result \cref{eq:localization_higher_derivs_gg}, so that localization now provides an independent check rather than an input. Motivated by the uniqueness of this solution and its agreement with localization, we conjecture that \cref{solC1234} gives the physical form factor. The corresponding worldsheet integrand is
    \begin{equation}
        \begin{aligned}
            \vec{P}_3^3(S,Q;x) = {}&\Big(\frac{1}{2} S^2 (2Q+S),-\frac{3}{2} S (2Q+S)^2,-\frac{3}{2} S (2 Q+S)^2,-4 Q^2(2 Q+S),\\
            &-S \left(2 Q^2+3 QS+S^2\right),-2 Q \left(6 Q^2+5 QS+S^2\right),-2 Q \left(6 Q^2+5 QS+S^2\right),\\
            &8 Q^2 (2Q+S),0,0,\frac{1}{2} S^2 (2Q+S)\Big)\\
            \vec{P}_2^2(S,Q;x) = {}&\Big(S (2 Q+S),28 Q^2+22 Q S+3S^2,24 Q^2+16 Q S+S^2,\\
            &-8 Q (5 Q+3S),2 S (2 Q+S)\Big)\;.\\
            \vec{P}_1^1(S,Q;x) ={}& \Big(-12 S+4 Q (x-3),2 (10 Q+S (x+9))\Big)\;,\\
            \vec{P}_0^0(S,Q;x) = {}&\Big(16\Big)\;.
        \end{aligned}
    \end{equation}
    The CFT data is also much simpler with these values for the coefficients. For the spectrum, we find:
    \begin{equation}
        \begin{aligned}
        \frac{\expval{\lambda^{(0)}_{n,2(n-2),\text{bulk,GG}}\widehat{\Delta}_{n,2(n-2),\text{bulk,GG}}^{(1)}}}{\expval{\lambda^{(0)}_{n,2(n-2),\text{bulk,GG}}}} ={}& \frac{3n^2+3n-2}{2\sqrt{n}}\;,\\
        \frac{\expval{\lambda^{(0)}_{n,2(n-3),\text{bulk,GG}}\widehat{\Delta}_{n,2(n-3),\text{bulk,GG}}^{(1)}}}{\expval{\lambda^{(0)}_{n,2(n-3),\text{bulk,GG}}}}={}& \frac{n(9n^2+37n+34)}{6\sqrt{n}(n+4)}\;,\\
        \frac{\expval{\lambda^{(0)}_{n,n-1,\text{def,GG}}\widehat{\Delta}_{n,n-1,\text{def,GG}}^{(1)}}}{\expval{\lambda^{(0)}_{n,n-1,\text{def,GG}}}} ={}& \frac{3n^2-3n+2}{4\sqrt{n}}\;,\\
        \frac{\expval{\lambda^{(0)}_{n,n-3,\text{def,GG}}\widehat{\Delta}_{n,n-3,\text{def,GG}}^{(1)}}}{\expval{\lambda^{(0)}_{n,n-3,\text{def,GG}}}} ={}& \frac{9n^3+13n^2-28n+24}{12\sqrt{n}(n+2)}\;.
        \end{aligned}
    \end{equation}
    Finally, the corrections to the OPE coefficients simplify as well. We have:
    \begin{equation}
        \begin{aligned}
        \expval{\lambda^{(1)}_{n,2(n-1),\text{bulk,GG}}} ={}& \frac{2^n n^n}{\Gamma(n)}\left(n^3 \zeta (3)-\frac{7 n^3}{12}+2 n^2\zeta(2)+\frac{7 n^2}{4}-\frac{13n}{6}+1\right)\;,\\
        \expval{\lambda^{(1)}_{n,n-1,\text{def,GG}}} ={}& \frac{3n^{n-3+q}2^{11-2n-5q}}{4(n-q+1)\Gamma(n)}\Bigg(n^4 (12 \zeta (3)-7)+n^3 \Big(-12 q \zeta (3)+25q+12 \zeta (3)\\
        &-12 \zeta(2)-40\Big)+n^2 \left(-31q^2+2 q (6 \zeta(2)+23)-12\zeta(2)-29\right)\\
        &+2 n \left(9 q^2-77q+98\right)+12 \left(11 q^2-27 q+16\right)\Bigg)\;.
        \end{aligned}
    \end{equation}

    \section{Discussion}\label{sec:discussion}
    In this paper, we initiated the study of AdS form factors describing strings scattering off D-branes. As concrete examples, we considered two-point functions of the superconformal primary of the stress tensor multiplet in 4d $\mathcal{N}=4$ SYM in the presence of either a $\frac12$-BPS 't Hooft line or a zero-dimensional defect formed by a pair of maximal giant graviton operators. These are dual to a D1 brane and a D3 brane, respectively. In both cases, we found explicit worldsheet representations for the leading curvature corrections to the AdS form factors. For the 't Hooft line, the form factor is uniquely fixed within the ansatz built from diagonal limits of SVMPLs, and the result passes nontrivial independent checks from supersymmetric localization. For the giant graviton, this restricted ansatz must be enlarged to general MPLs. The resulting worldsheet bootstrap leaves five coefficients unfixed. Localization fixes one of them, leaving a four-parameter family; alternatively, requiring that no new operators appear in the defect channel at order $1/R^2$ uniquely fixes the form factor without using localization, with the resulting answer reproducing the localization constraint. The leading curvature correction also encodes an infinite set of CFT data, some of which we presented for the first few Regge trajectories. There are many future directions based on our results and we will outline some of them below.

    First, the tension that appears in the giant graviton case between supersymmetric localization and the diagonal limit of single-valued multiple polylogarithms in the ansatz  is quite curious and worth further systematic investigation. In this paper, the two setups contain D-branes of different dimensions which might also be causing such a difference. It would therefore be useful to consider more examples of defects of the same D-brane origin, such as non-maximal giant gravitons and dual giant gravitons in $\mathcal{N}=4$ SYM, to allow for a less ambiguous comparison. These examples contain the same D3-brane, but the brane is embedded differently in AdS$_5\times$S$^5$. Computing their AdS form factors will help clarify whether the larger set of polylogarithmic functions in the maximal giant graviton integrand ansatz is needed in general. It is also interesting to point out that, for these examples, the higher order $1/\lambda$ corrections in the integrated correlators are independent of the giant graviton and dual giant graviton sizes \cite{Brown:2025huy}, which is a fact not at all obvious from the worldsheet perspective.

    In addition to considering correlators of two bulk local operators in the presence of a defect, we can also consider defect correlators where we have one operator in the bulk and two operators inserted on the defect. Kinematically, the latter type of defect correlators is not more complicated than the first kind. In a sense, these defect three-point functions also interpolate between defect two-point functions and four-point functions where all operators are on the defect. It is an interesting question to find out if the worldsheet approach continues to work for these defect correlators. 

    Finally, it would be interesting to pursue higher order curvature corrections and test the robustness of the worldsheet description assumption. However, more information is needed to bootstrap these corrections and other techniques, such as integrability, must first be developed to produce enough input. 

    \section*{Acknowledgements}
    We are grateful to Junding Chen for collaboration on an earlier version of this work. The work of CV is supported by the Fonds de recherche du Qu\'ebec, secteur Nature et technologies, through its Doctoral Training Scholarships program, scholarship number 349150 (\hyperlink{https://doi.org/10.69777/349150}{https://doi.org/10.69777/349150}), and the Natural Sciences and Engineering Research Council of Canada (NSERC), scholarship number CGRS D -- 610229 -- 2026. The work of X.Z. is supported by the NSFC Grant No. 12275273 and funds from Chinese Academy of Sciences. LFA is supported by the STFC grant ST/T000864/1.

    For the purpose of open access, the authors have applied a CC BY public copyright licence to any Author Accepted Manuscript (AAM) version arising from this submission.
    
    \appendix
    \section{Conformal blocks in Mellin space}\label{sec:conformal_blocks_mellin}
    In this appendix, we illustrate our approach to finding Mellin representations for the conformal blocks, in both the bulk and defect channels. In both cases, we define the conformal blocks as solutions to the conformal Casimir equation, but our approach for obtaining Mellin representations of these will be different.

    \subsection{Bulk channel}
    There is no known closed form for the Mack polynomials. Instead, we will use the fact that conformal blocks are eigenfunctions of conformal Casimir, which leads to the following differential equation for conformal blocks $g_{\Delta,\ell}^{\Delta_1,\Delta_2}(\xi,\chi)$~\cite{Billo:2016cpy}%
    \footnote{The added power of $\xi$ makes the Casimir operator look much simpler~\cite{Billo:2016cpy}, but Mellin space is best defined in our conventions, i.e. in terms of $g_{\Delta,\ell}^{\Delta_1,\Delta_2}(\xi,\chi)$.}:
    \begin{equation}
        \mathcal{C}_{\text{bulk}} \left(\xi^{(\Delta_1+\Delta_2)/2}g_{\Delta,\ell}^{\Delta_1,\Delta_2}(\xi,\chi)\right)=0\label{eq:casimir_equation}\;,
    \end{equation}
    where we defined the Casimir operator:
    \begin{equation}
        \begin{aligned}
        \mathcal{C}_{\text{bulk}}={}&2\xi^2(2\eta^2+\eta\xi+2)\pdv[2]{}{\xi}+2(\eta^2-1)(2\eta^2+2\eta\xi-2)\pdv[2]{}{\eta}\\
        &-4(1-\eta^2)\xi(2\eta+\xi)\pdv{}{\xi}{\eta}+2 \xi  \left(2 \eta ^2+\eta  \xi -2d+2\right)\pdv{}{\xi}\\
        &+\left(2 \xi  \left(d+\eta ^2-p-2\right)+4 \eta \left(\eta^2-1\right)\right)\pdv{}{\eta}-\frac{1}{2} \Delta_{12}^2 \eta  (2 \eta +\xi )+\Delta_{12}^2-2 C_{\Delta,\ell}\;.\label{eq:casimir_operator}
        \end{aligned}
    \end{equation}
    In this expression, $\Delta_{12}=\Delta_1-\Delta_2$, $p$ is the dimension of the defect, $d$ is the dimension of the CFT, and $C_{\Delta,\ell}$ is the eigenvalue of the Casimir operator
    \begin{equation}
        C_{\Delta,\ell} = \Delta(\Delta-d)+\ell(\ell+d-2)\;.
    \end{equation}
    In order to obtain exchange diagrams in Mellin space, we can translate this operator by first rewriting the action of the Casimir on $g_{\Delta,\ell}^{\Delta_1,\Delta_2}(\xi,\chi)$ using \cref{eq:casimir_equation}, and then acting on the Mellin transform \cref{eq:mellin_transform} to obtain a difference equation for the Mellin amplitude. Finally, we use \cref{eq:mellin_poles} to rewrite this as a difference relation for the Mack polynomials, which we can solve order-by-order in $\gamma$, starting with the leading order, $\gamma^{\ell/2}$. This fixes the blocks up to normalization constant $\mathcal{N}_{\Delta,\ell}^{\text{bulk}}$, which we will fix by demanding good flat space behaviour. In general, for the cases of interest in this paper we find:
    \begin{equation}
        \begin{gathered}
            Q_{\Delta,\ell,n}^{2,2,\text{line}}(\gamma) = \mathcal{N}_{\Delta,\ell}^{\text{bulk,line}}\left(\frac{2^{\Delta-1}\Gamma\left(\frac{\Delta-1}{2}+n\right)\Gamma\left(\frac{\Delta}{2}\right)\Gamma\left(\frac{\Delta-\ell}{2}+n-1\right)}{\sqrt{\pi}\Gamma(n+1)\Gamma(\Delta+n-1)\Gamma\left(\frac{\Delta-\ell}{2}\right)}\right)\gamma^{\frac{\ell}{2}}+O\left(\gamma^{\frac{\ell}{2}-1}\right)\\
            Q_{\Delta,\ell,n}^{4,4,\text{GG}}(\gamma) = \mathcal{N}_{\Delta,\ell}^{\text{bulk,GG}}\left(\frac{2^\Delta\Gamma\left(\frac{\Delta}{2}+n-1\right)\Gamma\left(\frac{\Delta+1}{2}\right)\Gamma\left(\frac{\Delta-\ell}{2}+n-3\right)}{\sqrt{\pi}\Gamma(n+1)\Gamma(\Delta+n-1)\Gamma\left(\frac{\Delta-\ell}{2}-2\right)}\right)\gamma^{\frac{\ell}{2}}+O\left(\gamma^{\frac{\ell}{2}-1}\right)
        \end{gathered}
    \end{equation}

    \subsection{Defect channel}
    In the defect channel, there are two conformal Casimir equations, and both must be obeyed~\cite{Billo:2016cpy}:
    \begin{equation}
        \mathcal{C}^\text{def}_{L^2} f_{\text{def}}(\Delta,s) = 0,\quad \mathcal{C}^\text{def}_{S^2} f_{\text{def}}(\Delta,s) = 0\;,
    \end{equation}
    where we defined the two operators $\mathcal{C}^\text{def}_{L^2}$ and $\mathcal{C}^\text{def}_{S^2}$ by
    \begin{equation}
        \begin{gathered}
        \mathcal{C}^\text{def}_{L^2}=(4-y^2)\pdv[2]{}{y}-(p+1)y\pdv{}{y}+C^{\text{def}}_{\Delta,0}\;,\\
        \mathcal{C}^\text{def}_{S^2} = 4\eta(1-\eta)\pdv[2]{}{\eta}+2(1-q\eta)\pdv{}{\eta}+C^{\text{def}}_{0,s}\;,
        \end{gathered}
    \end{equation}
    where $C_{\Delta,s}^{\text{def}}=\Delta(\Delta-p)+s(s+q-2)$ and $p+q=d$. We also defined the cross-ratio $y=\xi+\chi$%
    \footnote{In~\cite{Billo:2016cpy}, $y$ was denoted by $\chi$, but we choose to use the definition of $\chi$ from~\cite{Alday:2024srr}.}.
    An exact solution to both of these exists~\cite{Billo:2016cpy}:
    \begin{equation}
        \begin{aligned}
            f_{\Delta,s}(y,\eta) ={}& y^{-\Delta}\, _2F_1\left(\frac{\Delta+1}{2},\frac{\Delta}{2};\Delta+1-\frac{p}{2};\frac{4}{y^2}\right)\\
            &\times\,_2F_1\left(\frac{q+s}{2}-1,-\frac{s}{2};\frac{q-1}{2};1-\eta^2\right)\;.
        \end{aligned}
    \end{equation}
    In order to rewrite this in Mellin space, we note that the spin dependence of the conformal blocks factorizes:
    \begin{equation}
        f_{\Delta,s} = P_s(\eta)f_{\Delta,0}\;,
    \end{equation}
    where we defined the spin-dependence:
    \begin{equation}
        P_s(\eta)=\,_2F_1\left(\frac{q+s}{2}-1,-\frac{s}{2};\frac{q-1}{2};1-\eta^2\right)\;.
    \end{equation}
    We note that $P_s(\eta)$ is a polynomial of degree $s$, where all terms have the same parity, i.e. when $s$ is even, only even powers appear, and when $s$ is odd, only odd powers appear. In \cref{sec:scwi}, we show that, in Mellin space, multiplication by $\chi^n = \left(2\eta\right)^n$ acts as a shift in Mellin space:
    \begin{equation}
        \chi^n\circ \mathcal{M}(\delta,\gamma)= (\gamma-\delta)_n\prod_{i=1}^2\left(\frac{\Delta_i-\gamma}{2}\right)_{-\frac{n}{2}}\mathcal{M}(\delta,\gamma+n)\;.
    \end{equation}
    Due to the odd-argument Pochhammers above, it is best to separate between even and odd cases. However, given this result, we can write in Mellin space, for an individual conformal block:
    \begin{equation}
        M_{\Delta,s=2n}(\delta,\gamma) = \mathcal{N}_{\Delta,s}^{\text{def}}P_{2n}(\eta)\circ M_{\Delta,0}(\delta,\gamma)\;,\quad M_{\Delta,s=2n+1}(\delta,\gamma) = \mathcal{N}_{\Delta,s}^{\text{def}}\left(\frac{P_{2n+1}(\eta)}{\eta}\right)\circ M_{\Delta,1}(\delta,\gamma)\;,\label{eq:defect_channel_higher_spin}
    \end{equation}
    where, as in the bulk channel, we inserted arbitrary normalization which we will fix when taking the flat-space limit. Thus, we have reduced the problem of finding the conformal blocks in Mellin space to that of finding expressions for the conformal blocks with spins $0$ and $1$. We can do this by directly Mellin transforming those blocks, but the answer was already found in~\cite{Gimenez-Grau:2023fcy}, as the Witten exchange diagrams for those spins%
    \footnote{In the defect-free case, the Witten exchange diagrams are generally in one-to-one correspondence with the conformal blocks. In our case, at least in the defect channel, however, we will find that, in general, the diagram for an exchange of an operator of spin $\ell$ is a linear combination of conformal blocks of spin $j\le\ell$.}.
    The Mack polynomial, as defined in \cref{eq:mellin_poles}, for spin 0 is given by
    \begin{equation}
        P_{\Delta,0,n}^{\Delta_1,\Delta_2}(\delta,\gamma) = -\frac{\pi^{p/2}\Gamma\left(\frac{\Delta+\Delta_1-p}{2}\right)\Gamma\left(\frac{\Delta+\Delta_2-p}{2}\right)}{8\Gamma\left(\Delta_1\right)\Gamma\left(\Delta_2\right)}\frac{\left(1+\frac{\Delta-\Delta_1}{2}\right)_n\left(1+\frac{\Delta-\Delta_2}{2}\right)_n}{n!\Gamma\left(\Delta+n+1-\frac{p}{2}\right)}\;,
    \end{equation}
    while the spin-1 solution is
    \begin{equation}
        P^{\Delta_1,\Delta_2}_{\Delta,1,n}(\delta,\gamma) = (\gamma-\delta)P^{\Delta_1+1,\Delta_2+1}_{\Delta,0,n}(\delta,\gamma)\;.
    \end{equation}
    From these, we can then build conformal blocks in the defect channel in Mellin space at any spin using \cref{eq:defect_channel_higher_spin}.

    \section{Flat-space limit of conformal blocks}\label{sec:flat_blocks}
    In this appendix, we show how to compute the flat-space limit of individual conformal blocks, and give results for these limits in both channels for line defect and giant gravitons. First, we'll introduce the method we use. To be explicit, let's consider a bulk channel exchange as in \cref{eq:mellin_poles}:
    \begin{equation}
        M_{\Delta,\ell}^{\Delta_1,\Delta_2,\text{bulk}}(\delta,\gamma) = \sum_{n=0}^\infty \frac{Q_{\Delta,\ell,n}^{\Delta_1,\Delta_2}(\gamma)}{\delta-\frac{\Delta_1+\Delta_2-\tau-2n}{2}}\;.
    \end{equation}
    We choose to put the normalization factors of the flat-space limit inside the $\mathcal{N}$ normalization factors of the Mack polynomials, such that \cref{eq:flat_space_general} becomes
    \begin{equation}
        \mathcal{M}(\delta,\gamma) = \int_0^\infty \dd\beta\, \beta^{\frac{1}{2}(\Delta_1+\Delta_2-p)-1}e^{-\beta}\mathcal{A}\left(S = -\frac{2\delta\beta}{R^2},Q=\frac{2\gamma\beta}{R^2}\right),\quad R\to\infty\;,\label{eq:flat_space_no_norm}
    \end{equation}
    We note that, in this equation, $\delta,\gamma$ scale as $R^2$. Furthermore, we note that $\Delta\sim R$ by \cref{eq:ads_delta_m}. The last piece of information we need is that the sum above is dominated by terms of order $n\sim \Delta^2\sim R^2$~\cite{Fitzpatrick:2011hu}. Then, we can perform the following rescalings:
    \begin{equation}
        \delta\to\widehat\delta R^2\;,\quad \gamma\to \widehat\gamma R^2\;,\quad \Delta\to\widehat{\Delta}R\;,\quad n\to xR^2\;.
    \end{equation}
    Following this rescaling, we can expand the summands in $R$, and replace the sum by an integral over $x$. Performing this integral gives us the large-$R$ limit of $M_{\Delta,\ell}^{\Delta_1,\Delta_2,\text{bulk}}(\delta,\gamma)$. Then, we need to perform the Borel transform. In general, it is easiest to do this with the integral transform of \cref{eq:flat_space_no_norm}, rather than its inverse transform. Thus, we assume a form factor building block of the form
    \begin{equation}
        A^{\Delta_1,\Delta_2}_{\widehat\Delta,\ell,\text{bulk}}(S,Q) = \sum_{k=0,\frac{1}{2},1,\ldots}^\infty \frac{1}{R^{2k}}\frac{A^{(k)}_{\text{res}}(S,Q)}{(S-\widehat{\Delta}^2)^{p(k)}}\;.\label{eq:flat_block_ansatz}
    \end{equation}
    We find that, for integer $k$, $p(k)=3k+1$ is sufficient, and, for half-integer values, the pole is of lower order than the next integer value. Hence, we set $p(k)=3\lceil k\rceil+1$. In general, $A^{(k)}_{\text{res}}(S,Q)$ is a polynomial in $S,Q$. Then, we plug \cref{eq:flat_block_ansatz} in \cref{eq:flat_space_no_norm}, and compare the resulting expressions with the flat-space limit of $Q_{\Delta,\ell,n}^{\Delta_1,\Delta_2}(\gamma)$. This fully fixes the residues $A^{(k)}_{\text{res}}(S,Q)$. The location of these poles must match those of the gamma functions in \cref{eq:flat_space_form_factor}, which justifies \cref{eq:flat_spectrum}. Expanding $\widehat\Delta$ as in \cref{eq:Delta_hat_R_series} and demanding that the full series only have $\frac{1}{R^{2k}}$ terms for integer $k$ (i.e. demanding $A_\text{res}^{(k)}(S,Q)$ vanishes unless $k$ is an integer), as we expect for these form factors, gives us \cref{eq:delta_half}.

    After performing this second expansion, we find that we are able to find closed forms for the first two orders of the conformal blocks expanded near flat space. The most convenient way to write these is to re-organize our expansion into a sum of poles:
    \begin{equation}
        A_{\widehat\Delta,\ell,\text{bulk}}^{\Delta_1,\Delta_2}(S,Q) = \sum_{k=0}^\infty \frac{1}{R^{2k}} \left(\sum_{j=1}^{3k+1}\frac{A_{j,\text{bulk}}^{(k)}(Q)}{(S-2n)^j}+\ldots\right)\;,
    \end{equation}
    where we put all regular contributions in $\ldots$. In the defect channel, we have a similar expression:
    \begin{equation}
        A_{\widehat\Delta,\ell,\text{def}}^{\Delta_1,\Delta_2}(S,Q) = \sum_{k=0}^\infty \frac{1}{R^{2k}} \left(\sum_{j=1}^{3k+1}\frac{A_{j,\text{def}}^{(k)}(S)}{(Q-n)^j}+\ldots\right)\;.
    \end{equation}
    For the purpose of matching to the worldsheet, it is sufficient to extract these divergences, such that, in practice, we set the regular parts to 0. We find that the residues of these poles have explicit expressions in terms of Jacobi and Gegenbauer polynomials, as well as their derivatives. Concretely, we find that the normalizations we need to insert are:
    \begin{equation}
        \begin{gathered}
        \mathcal{N}_{\Delta,\ell}^{\text{bulk,line}} = R^{3-\frac{\ell}{2}}2^{-\Delta}\left(1+\frac{\ell(\ell-2)}{4\Delta} + O\left(\Delta^{-3}\right)\right)\;,\\
        \mathcal{N}^{\text{bulk,GG}}_{\Delta,\ell}=R^{8}2^{-\Delta-1}\frac{\sqrt{\pi}\Gamma\left(\frac{\Delta-\ell}{2}-2\right)}{\Gamma\left(\frac{\Delta+1}{2}\right)}\;,\quad \mathcal{N}^{\text{def,line}}_{\Delta,\ell} =\mathcal{N}^{\text{def,GG}}_{\Delta,\ell}= R^{2(1-\bar q)}\;,
        \end{gathered}
    \end{equation}
    where we defined $\bar q$ as the parity of $\ell$, i.e.:
    \begin{equation}
        \ell\equiv \bar q \mod 2\;.
    \end{equation}
    We define $x = 1+\frac{2Q}{n}$ and $y=1+\frac{S}{n}$. We also define $C_{n}^{(a)}(x)$, the Gegenbauer polynomial, and $P^{(a,b)}_n(x)$, the Jacobi polynomial. In terms of these, define the shorthands:
    \begin{equation}
        \mathcal{C}_n(x) = C^{(1/2)}_{n}(x)\;,\quad \mathcal{P}_n(x) = P^{(-1/2,1/2)}_n(x)\;,\quad \mathcal{U}_{n}(x) = C^{(1)}_{n}(x)
    \end{equation}
    Then, we can write the flat space limit of the conformal blocks, beginning with the line defect case in the bulk channel:
    \begin{equation}
        \begin{aligned}
        A^{(0)}_{1,\text{bulk,line}}(x) ={}& -\frac{2^{-\frac{3\ell}{2}}n^{\frac{\ell-2}{4}}\Gamma\left(1+\frac{\ell}{2}\right)}{\Gamma\left(\frac{1+\ell}{2}\right)}\mathcal{C}_{\ell/2}(x)\;,\\
        A^{(1)}_{4,\text{bulk,line}}(x) ={}&  -\frac{2^{3-\frac{3\ell}{2}}n^{\frac{6+\ell}{4}}\Gamma\left(1+\frac{\ell}{2}\right)}{\Gamma\left(\frac{1+\ell}{2}\right)}\mathcal{C}_{\ell/2}(x)\;,\\
        A^{(1)}_{3,\text{bulk,line}}(x) ={}& -\frac{2^{3-\frac{3\ell}{2}}n^{\frac{2+\ell}{4}}\Gamma\left(1+\frac{\ell}{2}\right)}{\Gamma\left(\frac{1+\ell}{2}\right)}\left(\frac{5}{4}\mathcal{C}_{\ell/2}(x)+\left(2+\frac{Q}{n}\right)\partial_x\mathcal{C}_{\ell/2}(x)\right)\;,\\
        A^{(1)}_{2,\text{bulk,line}}(x) ={}&-\frac{2^{3-\frac{3\ell}{2}}n^{\frac{\ell-2}{4}}\Gamma\left(1+\frac{\ell}{2}\right)}{\Gamma\left(\frac{1+\ell}{2}\right)}\Bigg(\frac{2-2\ell-\ell^2+4\sqrt{n}\widehat{\Delta}^{(1)}_{n,\ell}}{16}\mathcal{C}_{\ell/2}(x)+2\partial_x\mathcal{C}_{\ell/2}(x)\Bigg)\;,\\
        A^{(1)}_{1,\text{bulk,line}}(x) ={}& -\frac{2^{3-\frac{3\ell}{2}}n^{\frac{\ell-14}{4}}\Gamma\left(1+\frac{\ell}{2}\right)}{384\Gamma\left(\frac{1+\ell}{2}\right)}\Bigg(n^2\left(-5\ell^2-24\sqrt{n}\widehat{\Delta}^{(1)}_{n,\ell}+4\ell\left(5+3\sqrt{n}\widehat{\Delta}^{(1)}_{n,\ell}\right)\right)\mathcal{C}_{\ell/2}(x)\\
        &-n\left(-2Q(\ell^2+18\ell-12)+n (96+6\ell+11\ell^2)+96\sqrt{n}Q\widehat{\Delta}^{(1)}_{n,\ell}\right)\partial_x\mathcal{C}_{\ell/2}(x)\\
        &+2(64n^2+Q^2(10-3\ell)\ell+nQ(-3\ell^2+10\ell-192))\partial_x^2\mathcal{C}_{\ell/2}(x)+128nQ\partial_x^3\mathcal{C}_{\ell/2}(x)\Bigg)\;.
        \end{aligned}
    \end{equation}
    Next, we consider the giant gravitons case's bulk channel:
    \begin{equation}
        \begin{aligned}
        A^{(0)}_{1,\text{bulk,GG}} ={}& -\frac{2^{-\frac{3\ell}{2}}\sqrt{\pi}\Gamma\left(1+\frac{\ell}{2}\right)}{n^3\Gamma\left(\frac{1+\ell}{2}\right)}\mathcal{P}_{\ell/2}(x)\;,\\
        A^{(1)}_{4,\text{bulk,GG}} ={}& -\frac{2^{3-\frac{3\ell}{2}}\sqrt{\pi}\Gamma\left(1+\frac{\ell}{2}\right)}{n\Gamma\left(\frac{1+\ell}{2}\right)}\mathcal{P}_{\ell/2}(x)\;,\\
        A^{(1)}_{3,\text{bulk,GG}} ={}& -\frac{2^{3-\frac{3\ell}{2}}\sqrt{\pi}\Gamma\left(1+\frac{\ell}{2}\right)}{n^3\Gamma\left(\frac{1+\ell}{2}\right)}\left(2n\mathcal{P}_{\ell/2}(x)+(2n+Q)\partial_x\mathcal{P}_{\ell/2}(x)\right)\;,\\
        A^{(1)}_{2,\text{bulk,GG}} ={}& -\frac{2^{3-\frac{3\ell}{2}}\sqrt{\pi}\Gamma\left(1+\frac{\ell}{2}\right)}{16n^4\Gamma\left(\frac{1+\ell}{2}\right)}\left(-n\left(24+2\ell+\ell^2-4\sqrt{n}\widehat{\Delta}^{(1)}_{n,\ell}\right)\mathcal{P}_{\ell/2}(x)+8(n-4Q)\partial_x\mathcal{P}_{\ell/2}(x)\right)\;,\\
        A^{(1)}_{1,\text{bulk,GG}} = {}&-\frac{2^{3-\frac{3\ell}{2}}\sqrt{\pi}\Gamma\left(1+\frac{\ell}{2}\right)}{192n^7\Gamma\left(\frac{1+\ell}{2}\right)}\Bigg(3n^3\left(40+\ell-24\sqrt{n}\widehat{\Delta}^{(1)}_{n,\ell}\right)\mathcal{P}_{\ell/2}(x)\\
        &-12n^2\left(31n-35Q+4\sqrt{n}Q\widehat{\Delta}^{(1)}_{n,\ell}\right)\partial_x\mathcal{P}_{\ell/2}(x)-48n(5n-7Q)Q\partial_x^2\mathcal{P}_{\ell/2}(x)+64Q^3\partial_x^3\mathcal{P}_{\ell/2}(x)\Bigg)\;.
        \end{aligned}
    \end{equation}
    We now consider the line defect in the defect channel, where we find:
    \begin{equation}
        \begin{aligned}
            A^{(0)}_{1,\text{def,line}}(y) ={}&- \frac{\sqrt{\pi}}{4}\left(\frac{n}{8}\right)^{\bar q} \mathcal{C}_{\ell}(y)\;,\\
            A^{(1)}_{4,\text{def,line}}(y) ={}&-\sqrt{\pi}n^2\left(\frac{n}{8}\right)^{\bar q}\mathcal{C}_{\ell}(y)\;,\\
            A^{(1)}_{3,\text{def,line}}(y) ={}&-\sqrt{\pi}n\left(\frac{n}{8}\right)^{\bar q}\partial_y\mathcal{C}_{\ell}(y)\;,\\
            A^{(1)}_{2,\text{def,line}}(y) ={}&-\frac{\sqrt{\pi}}{16n^2}\left(\frac{n}{8}\right)^{\bar q}\left(\left(8\sqrt{n}\widehat{\Delta}^{(1)}_{n,\ell}-1\right)n^2\mathcal{C}_{\ell}(y)-4S(S+3n)\partial_y^2\mathcal{C}_{\ell}(y)\right)\;,\\
            A^{(1)}_{1,\text{def,line}}(y) ={}&-\frac{\sqrt{\pi}}{48n^4}\left(\frac{n}{8}\right)^{\bar q}\Bigg(3n^3\bar q\left(8\sqrt{n}\widehat{\Delta}^{(1)}_{n,\ell}-3\right)\mathcal{C}_{\ell}(y) + 3n^2\left(2n+3S-8\sqrt{n}S\widehat{\Delta}^{(1)}_{n,\ell}\right)\partial_y\mathcal{C}_{\ell}(y)\\
            &+6nS(5n+3S)\partial_y^2\mathcal{C}_{\ell}(y)+4S^2(3n+S)\partial_y^3\mathcal{C}_{\ell}(y)\Bigg)\;.
        \end{aligned}
    \end{equation}
    Finally, we consider the defect channel for the giant gravitons:
    \begin{equation}
        \begin{aligned}
            A^{(0)}_{1,\text{def,GG}}(y) ={}& -\frac{2^{-4-5\bar q}n^{\bar q}}{9(\ell+1)}\mathcal{U}_{\ell}(y)\;,\\
            A^{(1)}_{4,\text{def,GG}}(y) ={}& -\frac{2^{-2-5\bar q}n^{2+\bar q}}{9(\ell+1)}\mathcal{U}_{\ell}(y)\;,\\
            A^{(1)}_{3,\text{def,GG}}(y) ={}& -\frac{2^{-2-5\bar q}n^{1+\bar q}}{9(\ell+1)}\left(-\mathcal{U}_{\ell}(y)+\partial_y\mathcal{U}_{\ell}(y)\right)\;,\\
            A^{(1)}_{2,\text{def,GG}}(y) ={}& -\frac{2^{-4-5\bar q}n^{-2+\bar q}}{9(\ell+1)}\left(2n^{5/2}\widehat{\Delta}^{(1)}_{n,\ell}\mathcal{U}_{\ell}(y)-3n(3n+S)\partial_y\mathcal{U}_\ell(y)-S(3n+S)\partial_y^2\mathcal{U}_{\ell}(y)\right)\;,\\
            A^{(1)}_{1,\text{def,GG}}(y) ={}& -\frac{2^{-4-5\bar q}n^{-4+\bar q}}{27(\ell+1)}\Bigg(6n^3\bar q\left(\sqrt{n}\widehat{\Delta}^{(1)}_{n,\ell}-6\right)\mathcal{U}_{\ell}(y)+3n^2\left(15n+12S-2\sqrt{n}S\widehat{\Delta}^{(1)}_{n,\ell}\right)\partial_y\mathcal{U}_{\ell}(y)\\
            &+3nS(9n+4S)\partial_y^2\mathcal{U}_{\ell}(y)+S^2(3n+S)\partial_y^3\mathcal{U}_{\ell}(y)\Bigg)\;.
        \end{aligned}
    \end{equation}

    \section{Superconformal Ward identities in flat space}\label{sec:scwi}
    In this appendix, we recap the SCWI in Mellin space for the line defect in $\mathcal{N}=4$ SYM, as well as their solutions. We furthermore show how to expand them in a small-curvature series.
    \subsection{Superconformal Ward identities in Mellin space}\label{sec:scwi_mellin}
    To translate the Ward identities \cref{eq:scwi} to Mellin space, we follow~\cite{Gimenez-Grau:2023fcy,Alday:2024srr}. First, define:
    \begin{equation}
        \mathbb{W}(z,\bar z) = \eval{\left(\partial_z+\frac{1}{2}\partial_\alpha\right)\mathcal{F}(z,\bar z,\alpha)}_{\alpha=z}\;.
    \end{equation}
    The SCWI then become $\mathbb{W}(z,\bar z)=0$. We now separate this into four equations, in two steps. First, define:
    \begin{equation}
        \mathbb{W}_+(z,\bar z) = \mathbb{W}(z,\bar z)+\mathbb{W}(1/z,1/\bar z),\quad \mathbb{W}_-(z,\bar z) = \frac{1-z^2}{2z}(\mathbb{W}(z,\bar z)-\mathbb{W}(1/z,1/\bar z))\;,
    \end{equation}
    which we then use to create four independent linear combinations:
    \begin{equation}
        \begin{aligned}
            \mathbb{W}_{++}(z,\bar z) = \mathbb{W}_+(z,\bar z)+\mathbb{W}_+(\bar z,z),\quad \mathbb{W}_{+-}(z,\bar z) = \frac{z\bar z}{(\bar z-z)(1-z\bar z)}(\mathbb{W}_+(z,\bar z) -\mathbb{W}_+(\bar z,z))\;,\\
            \mathbb{W}_{-+}(z,\bar z) = \mathbb{W}_-(z,\bar z)+\mathbb{W}_-(\bar z,z),\quad \mathbb{W}_{--}(z,\bar z) = \frac{z\bar z}{(\bar z-z)(1-z\bar z)}(\mathbb{W}_-(z,\bar z) -\mathbb{W}_-(\bar z,z))\;.
        \end{aligned}
    \end{equation}
    Now, all of these vanish by the SCWI. To translate the four resulting equations into Mellin space, we start by rewriting the derivatives in terms of the cross-ratios $\xi,\chi$. In terms of the $\mathcal{F}_i(\xi,\chi)$ defined in \cref{eq:F_sigma_decomposition}, the $\mathbb{W}_{\pm}$ are given by:
    \begin{align}
        \begin{split}
            \mathbb{W}_{+}(z,\bar{z})={}&\sum_{j=0}^{\Delta_m}\frac{(1-z)^{2j}}{z^j}\bigg(-\left(\frac{(1-z)^2}{2z}+2\right)\xi\partial_\xi+\left(\xi-\frac{(1-z)^2}{2z}\chi\right)\partial_\chi\\
            {}&\quad\quad-\frac{j(1-z)^2}{2z}-2j\bigg)\frac{\mathcal{F}_j(\xi,\chi)}{(-2)^j}\;,\\
            \mathbb{W}_{-}(z,\bar{z})={}&\sum_{j=0}^{\Delta_m}\frac{(1-z)^{2j}}{z^j}\bigg(-\left(\left(\frac{(1-z)^2}{2z}\right)^2+\frac{3(1-z)^2}{2z}+2\right)\xi\partial_\xi\\
            {}&\quad\quad+\left(\left(\frac{(1-z)^2}{2z}+1\right)\xi-\left(\left(\frac{(1-z)^2}{2z}\right)^2+\frac{(1-z)^2}{2z}\right)\chi\right)\partial_\chi\\
            {}&\quad\quad-j\left(\frac{(1-z)^2}{2z}\right)^2-\frac{3j(1-z)^2}{2z}-2j\bigg)\frac{\mathcal{F}_j(\xi,\chi)}{(-2)^j}\;,
        \end{split}
    \end{align}
    where $\Delta_m=\min\{\Delta_1,\Delta_2\}$. From these, we form the $\mathbb{W}_{\pm\pm}$, which can be written in terms of new combination of the cross-ratios, $\zeta_{\pm}^{(j)}$, which we define as~\cite{Alday:2024srr}:
    \begin{equation}
        \zeta^{(j)}_+=\frac{(1-z)^{2j}}{2z^j}+\frac{(1-\bar{z})^{2j}}{2\bar{z}^j}\;,\quad \zeta^{(j)}_-=\frac{z\bar{z}}{(\bar{z}-z)(1-z\bar{z})}\left(\frac{(1-z)^{2j}}{2z^j}-\frac{(1-\bar{z})^{2j}}{2\bar{z}^j}\right)\;.
    \end{equation}
    Finally, these combinations can be rewritten as polynomials in the $\xi,\chi$ variables~\cite{Gimenez-Grau:2023fcy}:
    \begin{equation}
        \begin{aligned}
            \zeta_{+}^{(i)}
            & = \sum _{j=0}^{\left\lfloor i/2\right\rfloor } \binom{i}{2 j}
            \left(\frac{\chi^2}{4}-1\right)^j \left((\chi +\xi )^2-4\right)^j
            \left(\frac{\chi^2}{2} +\frac{\xi\chi}{2}-2\right)^{i-2 j}\;,\\
            \zeta_{-}^{(i)}
            & = \sum _{j=0}^{\left\lfloor i/2\right\rfloor } \binom{i}{2 j+1}
            \left(\frac{\chi^2}{4}-1\right)^j \left((\chi+\xi )^2-4\right)^j
            \left(\frac{\chi^2}{2} +\frac{\xi\chi}{2}-2\right)^{i-2 j-1}\;.
        \end{aligned}
    \end{equation}
    Now, we are ready to translate the SCWI into Mellin space. From the Mellin transform \cref{eq:mellin_transform}, we can establish the following dictionary between operators on the cross-ratios and operators on the Mellin amplitude:
    \begin{equation}
        \begin{aligned}
            \xi\partial_\xi\to (-\delta)\;,\quad& \chi\partial_\chi\to (-\gamma+\delta)\;,\\
            \xi^m\chi^n\to (\delta)_m(\gamma-\delta)_n&\prod_{i=1}^2\left(\frac{\Delta_i-\gamma}{2}\right)_{-\frac{m+n}{2}}P_{m,m+n}\;,\label{eq:pos_to_mellin_defect}
        \end{aligned}
    \end{equation}
    where we defined the shift operators $P_{m,n}$:
    \begin{equation}
        P_{m,n}f(\delta,\gamma) = f(\delta+m,\gamma+n)\;.
    \end{equation}
    Using this dictionary, we obtain four independent SCWI equations, which take the form of difference equations between the $\mathcal{M}_i(\delta,\gamma)$ of \cref{eq:mellin_sigma_decomposition}, by setting $\mathbb{W}_{\pm\pm}=0$.

    In~\cite{Virally:2025nnl}, it was shown that these equations admit solutions in the form of a reduced correlator, acted upon by a difference operator. That is to say, there exists a solution of the SCWI of the form \cref{eq:reduced_definition}:
    \begin{equation}
        \mathcal{M}(\delta,\gamma;\sigma) = \mathcal{D}\widetilde{\mathcal{M}}(\delta,\gamma)
    \end{equation}
    where $\mathcal{D}$ is a difference operator which contains all of the dependence on $\sigma$. We refer readers interested in the method for deriving $\mathcal{D}$, which can be applied to a range of superconformal theories, to~\cite{Virally:2025nnl}. Here, we will simply quote the result. We can decompose $\mathcal{D}$ into $\sigma$-power components:
    \begin{equation}
        \mathcal{D} = \sum_{i=0}^2\sigma^i\mathcal{D}_i\;.
    \end{equation}
    Then, the solution to the SCWI in Mellin space is given by:
    \begin{equation}
        \begin{aligned}
        \mathcal{D}_0 ={}& -(\gamma -2) (\delta +1) P_{2,2}+(\delta +1) (\gamma
        -\delta +1)P_{2,4}+\frac{1}{2} (\gamma -2) \delta 
        P_{3,2}+\frac{1}{2}
        \delta  (-\gamma +2 \delta +1)
        P_{3,4}\;,\\
        \mathcal{D}_1 ={}&\frac{(\gamma -2) \left(2 \gamma ^2-6
        \gamma  \delta +3 \gamma +3 \delta
        ^2+\delta -4\right)}{\delta
        -1}P_{1,2}-\frac{(\gamma -\delta )
        \left(\gamma ^2-5 \gamma  \delta +7
        \gamma +4 \delta ^2-12 \delta
        +8\right)}{\delta -1}P_{1,4}\\
        &+\frac{2 (\gamma -4)(\gamma -2)^2}{\delta -1}P_{0,0}-\frac{(\gamma -4)(\gamma -2)^2 }{\delta -1}P_{1,0}-\frac{2 (\gamma -2)(\gamma -\delta +1) (2 \gamma -\delta-2)}{\delta-1}P_{0,2}\\
        &+(\gamma-2) (\gamma -\delta +1)
        P_{2,2}+\frac{2(\gamma -\delta ) (\gamma -\delta +1)(\gamma -\delta +3)}{\delta -1}P_{0,4}-(\gamma -2 \delta +2)(\gamma -\delta )P_{2,4}\;,\\
        \mathcal{D}_2 ={}& -\frac{(\gamma -4) (\gamma -2)^2}{\delta-1}P_{0,0}+\frac{(\gamma -4) (\gamma -2)^2}{2 (\delta-1)}P_{1,0}+\frac{(\gamma -3) (\gamma -2)(\gamma -\delta +1)}{\delta -1}P_{0,2}\\
        &-\frac{(\gamma -3)(\gamma -2) (\gamma -2 \delta +2)}{2 (\delta-1)}P_{1,2}\;.\label{eq:scwi_sol_op}
        \end{aligned}
    \end{equation}
    It is simple to check that any $\mathcal{M}(s,t)$, put into \cref{eq:reduced_definition} using this form of $\mathcal{D}$, leads to a solution of the SCWI in Mellin space. We now want to present some known solutions of the SCWI, which we need for the localization check in \cref{sec:line_localization}. The first of these is the SUGRA amplitude, which we already gave as a full amplitude in \cref{eq:line_sugra}, and in reduced form in \cref{eq:line_sugra_reduced}. The next few are polynomial solutions to the SCWI, which are used to build the higher-derivative corrections to SUGRA in \cref{eq:localization_higher_derivs}. These polynomial solutions were all found in~\cite{Alday:2024srr}. There are no constant or linear solutions to the SCWI, with the first solution being degree $L=2$ in $\delta,\gamma$, and given by:
    \begin{equation}
        M_{L=2}^{(1)}=\frac{1}{5} \left(5 \delta ^2+11 \delta +12-2 \sigma  (5 \gamma  \delta -32 \gamma +8 \delta +28)+(\gamma -2) (5 \gamma -14) \sigma ^2\right)\;.\label{eq:scwi_sol_L_2}
    \end{equation}
    There are then two solutions at order $L=3$:
    \begin{equation}
        \begin{aligned}
            M^{(1)}_{L=3}={}& \frac{1}{21}\Big(21\gamma\delta^2 + 51\gamma\delta + 60\gamma + 8\delta^2 + 8\delta - 2\sigma\left(21\gamma^2\delta - 174\gamma^2 + 44\gamma\delta + 244\gamma + 8\delta - 152\right) \\
            & + (\gamma - 2)\left(21\gamma^2 - 88\gamma + 76\right)\sigma^2\Big)\;, \\
            M^{(2)}_{L=3}={}& \frac{1}{7}\Big(\delta(\delta + 1)(7\delta + 4)- 2\sigma\left(7\gamma\delta^2 - 69\gamma\delta + 100\gamma + 18\delta^2 + 14\delta - 80\right) \\
            & + (\gamma - 2)\sigma^2\left(7\gamma\delta - 16\gamma - 20\delta + 44\right)\Big)\;.\label{eq:scwi_sol_L_3}
        \end{aligned}
    \end{equation}
    We also need the quartic polynomials, of which there are three:
    \begin{equation}
        \begin{aligned}
            M^{(1)}_{L=4}={}& \frac{1}{135}\Big(15\gamma^2(9\delta^2 + 23\delta + 28)+ 72\gamma\delta(\delta + 1)- 28\delta^2(\delta + 1) \\
            &+ \sigma\left(-30\gamma^3(9\delta - 92)- 48\gamma^2(13\delta + 123)+ 8\gamma(7\delta^2 - 91\delta + 1028)+ 16(9\delta^2 + 5\delta - 242)\right) \\
            &+ (\gamma - 2)\sigma^2\left(135\gamma^3 - 798\gamma^2 - 4\gamma(7\delta - 377) + 80\delta - 984\right)\Big)\;, \\
            M^{(2)}_{L=4}={}& \frac{1}{9}\Big(\delta(\delta + 1)\left(\gamma(9\delta + 4) - 4\delta\right)+ (\gamma - 2)^2\sigma^2\left(3\gamma(3\delta - 8) - 28\delta + 76\right) \\
            &+ \sigma\left(-2\gamma^2(9\delta^2 - 109\delta + 196)- 8\gamma(5\delta^2 + 34\delta - 104)+ 24(\delta^2 + 7\delta - 24)\right)\Big)\;, \\
            M^{(3)}_{L=4}={}& \frac{1}{18}\Big(3\delta^2(6\delta^2 + 13\delta + 7)+ \sigma\left(-6\gamma(6\delta^3 - 73\delta^2 + 165\delta - 100)- 128\delta^3 + 92\delta^2 + 484\delta - 480\right) \\
            &+ \sigma^2\left(3\gamma^2(6\delta^2 - 21\delta + 16)+ \gamma(-88\delta^2 + 302\delta - 228)+ 8(13\delta^2 - 44\delta + 33)\right)\Big)\;.\label{eq:scwi_sol_L_4}
        \end{aligned}
    \end{equation}
    For completeness, we also present the reduced amplitudes for all of these, from~\cite{Virally:2025nnl}:
    \begin{equation}
        \begin{gathered}
            \widetilde{M}^{(1)}_{L=2} = \frac{2}{5}\delta,\quad \widetilde{M}^{(1)}_{L=3} = \frac{2}{21}\delta(3\gamma-8)\;,\quad\widetilde{M}^{(2)}_{L=3} = \frac{2}{7}(\delta-1)(\delta-2)\;,\\ \widetilde{M}^{(1)}_{L=4} =\frac{2}{135} \left(\delta  \left(15\gamma ^2-84 \gamma -4 \delta+116\right)-8\right)\;,\quad \widetilde{M}^{(2)}_{L=4} =\frac{2}{9} (\gamma -4)(\delta -2)(\delta -1)\;,
            \\ \widetilde{M}^{(3)}_{L=4} =\frac{1}{9} (\delta -2)(\delta -1)(2\delta -3)\;.
        \end{gathered}
    \end{equation}

    \subsection{Flat-space limit}\label{sec:scwi_flat_limit}
    We now want to rewrite the solutions to the SCWI in Mellin space, i.e. \cref{eq:scwi_sol_op}, in the flat-space limit. To do this, we use the flat-space limit formula, \cref{eq:flat_space_general}. The first step is to rescale the Mellin variables in the usual way, $\delta=R^2\hat\delta$ and $\gamma=R^2\hat\gamma$. Then, we divide $\mathcal{D}$ by $R^2$ in order to normalize its scaling in the flat-space limit, and we can expand the shift operators as derivatives, which gives for example:
    \begin{equation}
        \mathcal{D}_0 = -q\hat\delta + \hat\delta^2\partial_{\hat \delta} +\hat\delta\hat\gamma\partial_{\hat\gamma} + O\left(R^{-2}\right)\;,
    \end{equation}
    and similarly for the other components in $\sigma$. Here, we defined $q=\frac{p-\Delta_1-\Delta_2}{2}$ such that \cref{eq:flat_space_general} becomes:
    \begin{equation}
        \mathcal{M}(\hat\delta,\hat\gamma) =\kappa\int_0^\infty \dd\beta\, e^{-\beta}\beta^{-q-1}\mathcal{A}\left(S=-2\hat\delta\beta,Q=2\hat\gamma\beta\right)\;,
    \end{equation}
    for some normalization $\kappa$. This is inverted by:
    \begin{equation}
        \mathcal{A}(S,Q) = \bar{\kappa} \int_{k-i\infty}^{k+i\infty}\frac{\dd\alpha}{2\pi i}\,e^{\alpha}\alpha^q\mathcal{M}\left(\hat\delta = -\frac{S}{2\alpha},\hat\gamma=\frac{Q}{2\alpha}\right)\;.
    \end{equation}
    In order to translate differential operators on $(\hat\delta,\hat\gamma)$ into operators on $(S,Q)$, we will use an approach similar to that of~\cite{Alday:2024xpq}. Specifically, we will map local operators on $(\hat\delta,\hat\gamma)$ space, in the large-$R$ limit of Mellin space, to other local operators in ``Borel space'', which is to say the space of the $(S,Q)$ variables. The operators we introduce for this purpose are:
    \begin{equation}
        \begin{aligned}
            U_0 = \hat\delta\partial_{\hat\delta},\quad U_+=2q\hat\delta -{}& 2\hat\delta^2\partial_{\hat\delta}-2\hat\delta\hat\gamma\partial_{\hat\gamma},\quad U_- = \partial_{\hat\delta}\;,\\
            V_0 = \hat\gamma\partial_{\hat\gamma},\quad V_+ = -2q\hat\gamma +{}& 2\hat\gamma^2\partial_{\hat\gamma} + 2\hat\delta\hat\gamma\partial_{\hat\delta},\quad V_- = \partial_{\hat\gamma}\;,\\
            W = \hat\gamma\partial_{\hat\delta},\quad \overline{W}=\hat\delta\partial_{\hat\gamma},\ \ & \ \ K = \frac{1}{\hat\delta},\quad H = \frac{\hat\gamma}{\hat\delta}\;.\label{eq:mellin_operators}
        \end{aligned}
    \end{equation}
    In Borel space, they have the following action:
    \begin{equation}
        \begin{aligned}
            U_0 = S\partial_S,\quad U_+ = S,\quad U_- = 2q\partial_S&-2S\partial^2_S - 2Q\partial_S\partial_Q\;,\\
            V_0 = Q\partial_Q,\quad V_+ = Q,\quad V_- = -2q\partial_Q& + 2Q\partial_Q^2 + 2S\partial_S\partial_Q\;,\\
            W = -Q\partial_S,\quad \overline{W} = -S\partial_Q,\quad K = -2\partial_S +{}& \frac{2(q+1)}{S} - \frac{2Q}{S}\partial_Q\quad H = -\frac{Q}{S}\;.\label{eq:borel_operators}
        \end{aligned}
    \end{equation}
    Thus, rewriting the SCWI operators in Borel space becomes a question of rewriting their expansions in Mellin space in terms of the operators in \cref{eq:mellin_operators} and then using \cref{eq:borel_operators}. Doing this, we find the following expressions for the flat-space limits of the operators $\mathcal{D}_i$:
    \begin{equation}
        \begin{aligned}
        \mathcal{D}_0 ={}& -\frac{1}{2}U_+ + \frac{1}{R^2}\Bigg(\frac{5}{2}U_0^2+4U_0\overline{W}+U_0V_0+3V_0\overline{W}+2V_0+3\Bigg) + O\left(R^{-4}\right)\;,\\
        \mathcal{D}_1 ={}& -V_+ + \frac{1}{R^2}\Bigg(2KV_+V_0+4KV_+-2U_0^2+U_0W-10U_0V_0+4WV_0\\
        &-4U_0+12W-6V_0^2-4V_0-2\Bigg)+ O\left(R^{-4}\right)\;,\\
        \mathcal{D}_2 = {}&\frac{1}{2}HV_++ \frac{1}{R^2}\Bigg(HV_0^2-HWV_0+H^2V_0 - 6HV_0 - \frac{3}{2}HW - 7H + \frac{3}{2}H^2 \\
        &+ \frac{1}{2}W^2+2WV_0-5W\Bigg)+O\left(R^{-4}\right)\;.
        \end{aligned}
    \end{equation}
    We note that, for each component of these operators, their flat space expansion is led by an operator which becomes a simple multiplication in Borel space. This means that, in a $\frac{1}{R^2}$ series, the SCWI operators are always invertible term-by-term in Borel space. We use this to find the reduced forms of each correlator in flat space.
    
    \section{Single valued multiple polylogarithms}\label{sec:mpls}
    In this appendix, we review some known facts about multiple polylogarithms (MPLs), as well as the related single valued functions. We refer the reader to~\cite{BrownMPL2004,Dixon:2012yy} for a more detailed description, as well as to~\cite{Maitre:2005uu,Duhr:2019tlz} for both detailed descriptions of these functions and packages which implement them in \texttt{Mathematica}. MPLs, also known as harmonic polylogarithms (HPLs) in the literature, are functions of a single variable $z$, labelled by a word $w$ made up of letters in the alphabet $\{0,1\}$. We define these functions recursively by the differential relation:
    \begin{equation}
        \dv{}{z}L_{aw} = \frac{1}{z-a}L_w(z)\;,\label{eq:MPL_diff_relation}
    \end{equation}
    as well as the boundary condition $\lim_{z\to 0}L_w(z)=0$. There is a single exception to this boundary condition, which is when $w=0^p$ in which case we define $L_w(z) = \frac{\log^p(z)}{p!}$. For the empty word, we thus have $L_{\emptyset}(z)=1$. We call the length of the word, $|w|$, the transcendentality weight of $L_{w}(z)$. For certain words, $L_w(z)$ can be expressed in terms of other special functions, for example:
    \begin{equation}
        L_{0^{n-1}1}(z) = -\Li_n(z)\;.
    \end{equation}
    MPLs are closed under particular transformations of $z$, in particular:
    \begin{equation}
        z\to 1-z,\quad z\to \frac{z}{z-1}\;,
    \end{equation}
    by which we mean that $L_w(1-z)$ and $L_w(\frac{z}{z-1})$ can be rewritten in terms of linear combinations of $L_{w'}(z)$ with $|w'|\le |w|$ and multiple zeta values. These can be explicitly computed by using the nested integral structure implied by \cref{eq:MPL_diff_relation}:
    \begin{equation}
        L_{aw}(z) = \int_0^z\dd z'\, \frac{1}{z'-a}L_w(z')\;.
    \end{equation}
    Finally, we note that the $L_w(z)$ obey shuffle relations, which can be written as:
    \begin{equation}
        L_w(z)L_{w'}(z) = \sum_{W\in w\shuffle w'}L_W(z)\;,\label{eq:mpl_shuffle}
    \end{equation}
    where $\shuffle$ denotes the shuffle product, which is to say the sum of all ways to permute the letters in the union of two words while keeping their relative order in each word fixed. For example, with two words $ab$ and $cd$, the shuffle product would be~\cite{Duhr:2019tlz}:
    \begin{equation}
        ab\shuffle cd = abcd + acdb + cabd + acdb + cadb + cdab\;.
    \end{equation}
    MPLs are analytic functions with branch points at $z=0,1,\infty$ on the Riemann sphere. We define single valued multiple polylogarithms (SVMPLs) as:
    \begin{equation}
        \mathcal{L}_{W}(z) = \sum_{|w|+|w'|\le |W|} c_{w,w'}L_{w}(z)L_{w'}(\bar z)\;,
    \end{equation}
    where the $c_{w,w'}$ are chosen such that the branch cuts' discontinuities cancel. The SVMPLs are further defined to obey the same differential relation as the MPLs:
    \begin{equation}
        \pdv{}{z}\mathcal{L}_{aw} = \frac{1}{z-a}\mathcal{L}_w(z)\;,\label{eq:SVMPL_diff_relation}
    \end{equation}
    along with the same boundary condition $\lim_{z\to 0}\mathcal{L}_w(z)=0$, except for $\mathcal{L}_{0^p}(z) = \frac{\log^p z\bar z}{p!}$. They also obey the same shuffle relations:
    \begin{equation}
        \mathcal{L}_w(z)\mathcal{L}_{w'}(z) = \sum_{W\in w\shuffle w'}\mathcal{L}_W(z)\;.\label{eq:svmpl_shuffle}
    \end{equation}
    The explicit construction of SVMPLs in terms of MPLs can be found in~\cite{BrownMPL2004,Dixon:2012yy}, and is implemented in the package~\cite{Duhr:2019tlz}. In this work, we are primarily interested in the diagonal limit of SVMPLs, which is defined by:
    \begin{equation}
        \mathcal{L}^{\text{diag}}_w(x) = \eval{\mathcal{L}(z)}_{\bar z=z=x}\;.
    \end{equation}
    In practice, when working with the diagonal limit of the SVMPLs, we rewrite these as linear combinations of MPLs by using their explicit construction and the shuffle relations \cref{eq:mpl_shuffle}. The main gain of using the diagonal limits of SVMPLs, rather than generic linear combinations of MPLs, is then that this is a more restrictive ansatz, as not all MPL expressions can be obtained from combinations of diagonal limits of SVMPLs. In general, we have the following relationship~\cite{Alday:2024ksp}:
    \begin{equation}
        \mathcal{L}_w(z)+\mathcal{L}_w(\bar z) = \mathcal{L}_{\tilde{w}}(z)+\mathcal{L}_{\tilde{w}}(\bar z)\;,
    \end{equation}
    where $\tilde w$ is the reverse word of $w$. When we take the diagonal limit $\bar z=z=x$, this becomes:
    \begin{equation}
        \mathcal{L}^{\text{diag}}_{w}(x) = \mathcal{L}^{\text{diag}}_{\tilde w}(x)\;. 
    \end{equation}
    These relationships explain why the basis of MPLs in \cref{eq:mpl_basis} is slightly smaller than would be found by simply writing all possible words of length 3 or less made of letters $\{0,1\}$.

    \subsection{Integrating the worldsheet}
    We conclude this appendix by showing how we can integrate our MPL ansatz on the worldsheet. In general, recall that we are integrating terms of the form:
    \begin{equation}
        I=\int_0^1\dd x\, x^{-\frac{S}{2}-1}(1-x)^{-Q-1}L_w(x)x^\epsilon\;,
    \end{equation}
    where $\epsilon$ is $0$ or $1$, and comes from the linear $x$ terms in the $P_w^a(S,Q;x)$ in \cref{eq:integrand_ansatz}. Specifically, we want to extract individual poles in $S$ from this integral, as this is what we want to compare to CFT exchanges in practice. In order to extract poles in $Q$, we perform the change of variables $x\to1-x$ and apply the same logic. Poles in $S$ comes from the expansion of the integrand near $x=0$. Near this point, $(1-x)^{-Q-1}$ is analytic. Furthermore, $\mathcal{L}_w^{\text{diag}}(x)$ has an expansion in powers of $x$ and $\log x$:
    \begin{equation}
        L_w(x) = \sum_{\substack{0\le a,\\ 0\le b\le |w|}}\alpha_{a,b,w}x^a\log^bx\;.
    \end{equation}
    Then, we note that:
    \begin{equation}
        \int_0^1 \dd x\, x^{-\frac{S}{2}-1+a}\log^b x = (-2\partial_S)^b\int_0^1 x^{-\frac{S}{2}-1+a}\;.
    \end{equation}
    Finally, we note that this last integral is trivial:
    \begin{equation}
        \int_0^1 x^{-\frac{S}{2}-1+a} = \frac{2}{2a-S}\;.
    \end{equation}
    Putting everything in this appendix together, it is possible to perform all of the integrals on the worldsheet required to obtain poles of the form factor by expanding MPLs near $x=0$ or $x=1$.

    \section{Additional data in the giant graviton system}\label{sec:gg_CFT_data}
    In this appendix, we present some additional CFT data for the giant graviton system. First, we present the anomalous dimension in the first accessible Regge trajectory in the defect channel:
    \begin{equation}
        \begin{aligned}
            \frac{\expval{\lambda^{(0)}_{n,n-3,\text{def,GG}}\widehat{\Delta}_{n,n-3,\text{def,GG}}^{(1)}}}{\expval{\lambda^{(0)}_{n,n-3,\text{def,GG}}}} ={}& \frac{1}{10\sqrt{n}(n+2)}\Big(10n^3C_1+10(C_4+\zeta(2)(n+2)C_3)n^2\\
            &+(14C_4+46C_2-12C_1+5)n+2(5+16C_1-8C_2-12C_4)\Big)\;.
        \end{aligned}
    \end{equation}
    For the first Regge trajectories in the bulk channel after the Konishi trajectory, we find:
    \begin{equation}
        \begin{aligned}
        \frac{\expval{\lambda^{(0)}_{n,2(n-2),\text{bulk,GG}}\widehat{\Delta}_{n,2(n-2),\text{bulk,GG}}^{(1)}}}{\expval{\lambda^{(0)}_{n,2(n-2),\text{bulk,GG}}}} ={}& \frac{1}{20\sqrt{n}}\Big((60-106C_1-27C_2-100C_3+27C_4)n^2\\
        &+3(20+74C_1+83C_2+100C_3-3C_4)n\\
        &-2(40+58C_1+111C_2+100C_3+9C_4)\Big)\;,\\
        \frac{\expval{\lambda^{(0)}_{n,2(n-3),\text{bulk,GG}}\widehat{\Delta}_{n,2(n-3),\text{bulk,GG}}^{(1)}}}{\expval{\lambda^{(0)}_{n,2(n-3),\text{bulk,GG}}}} ={}&\frac{1}{60\sqrt{n}(n+4)}\Big(90n^3\\
        &+(597C_4+1060C_3-597C_2-246C_1-540)n^2\\
        &+(3000-228 C_1+2634 C_2-2600 C_3-474C_4)n\\
        &-48(6 C_1-3 C_2-20 C_3+3C_4-10)\Big)\;.
        \end{aligned}
    \end{equation}
    Next, there are the $\frac{1}{R^2}$ corrections to the OPE coefficients. In the bulk channel, we have:
    \begin{equation}
        \begin{aligned}
            \expval{\lambda_{n,2(n-1),\text{bulk,GG}}^{(1)}} ={}& \frac{2^nn^n}{\Gamma(n)}\Bigg(-\frac{2}{5} \left(4 C_1+3 C_2-3C_4\right) n^3 \zeta(3)\\
            &+\frac{1}{120} \left(-318 C_1-81C_2+60 C_3+81 C_4+20\right)n^3\\
            &+\frac{1}{10} n^2 \zeta(2)\left(2 \left(8 C_1+6 C_2-10 C_3-6C_4+5\right) n-4 C_1+7 C_2+20 C_3+3C_4+25\right)\\
            &+\frac{1}{8} \left(18C_1+11 C_2-4 C_3-3 C_4+12\right)n^2-\frac{13 n}{6}+1\Bigg)\;.
        \end{aligned}
    \end{equation}
    In the defect channel, we find:
    \begin{equation}
        \begin{aligned}
            \expval{\lambda_{n,n-1,\text{def,GG}}^{(1)}} ={}& \frac{n^{n-3+q}2^{11-2n-5q}}{\Gamma(n)}\Bigg(\frac{9}{10} \left(6 C_1-3 C_2+20 C_3+3 C_4\right)n^3 \zeta (3)\\
            &+\frac{9}{20} n^2 \zeta(2)\Big(20 C_3 (n+2 q-2)+C_1 (50 n-8)+C_2 (15n+14)+C_4 (6-15 n)\\
            &-10 (n+1)\Big)+18 n^2\Bigg(-\frac{C_2 \left(9 n^2-36 n-20q+40\right)}{20 n}-\frac{1}{5} C_1 (3 n-5q+3)\\
            &+\frac{C_4 (3 n-5 q+3) (3 n+2 q-2)}{20(n-q+1)}+C_3 n+\frac{-8 n^3+9 n^2-16 n-66 q+96}{12n^2}\Bigg)\Bigg)\;.
        \end{aligned}
    \end{equation}
    Finally, we consider the new operators in the defect channel of the giant graviton system. The first new OPE coefficient is:
    \begin{equation}
        \begin{aligned}
            \expval{\lambda_{n,n-2,\text{def,GG}}^{(1)}} = {}& \frac{2^{5-2n+5q}(n-1)^2n^{n-q-2}}{\Gamma(n)}\Bigg(-3 n \Big(3 \left(10 C_1+3 C_2+4 C_3-3C_4-2\right) \zeta(2)\\
            &+10 C_1-15C_2+78 C_3+15C_4-35\Big)-\frac{33}{5} \left(4C_1+33 C_2-3 C_4+25\right)\Bigg)\;.
        \end{aligned}
    \end{equation}

    \bibliographystyle{jhep.bst}
    \bibliography{AdS_Defects.bib}
\end{document}